\documentclass{aa}
\usepackage{natbib}
\bibpunct{(}{)}{;}{a}{}{,} % to follow the A&A style

\def\gr{$\gamma$-ray}
\usepackage[normalem]{ulem}
\usepackage[utf8]{inputenc}
\usepackage{graphicx}
\usepackage{hyperref}
\usepackage{color}
\graphicspath{{./figs/}}

\newcommand{\dd}{{\rm \, d}}
\newcommand{\cs}{c_{\rm s}}
\newcommand{\vA}{v_{\rm A}}

\begin{document}

\title{LISA and \gr\ telescopes as multi-messenger probes of a first-order cosmological phase transition}
\author{
A. Roper Pol\inst{\ref{inst2}}\fnmsep\thanks{corresponding author [\email{alberto.roperpol@unige.ch}]}
\and A. Neronov\inst{\ref{inst1},} \inst{\ref{inst4}} \and 
C. Caprini\inst{\ref{inst2},}\inst{\ref{inst3}} \and
T. Boyer\inst{\ref{inst1}}  \and D. Semikoz\inst{\ref{inst1}}}
\institute{D\'epartement de Physique Th\'eorique, Universit\'e de Gen\`eve,
CH-1211 Gen\`eve, Switzerland.\label{inst2}
\and Universit\'e Paris Cit\'e, CNRS, Astroparticule et Cosmologie, F-75006 Paris, France.\label{inst1}
\and Laboratory of Astrophysics, \'Ecole Polytechnique F\'ed\'erale de Lausanne,
CH-1015 Lausanne, Switzerland.\label{inst4}
\and Theoretical Physics Department, CERN, CH-1211 Gen\`eve, Switzerland.\label{inst3}}

\titlerunning{SGWB and IGMF}
\authorrunning{}
\abstract
{}
{
We study two possible cosmological consequences of a first-order phase transition in
the temperature range of $1$~GeV to $10^3$~TeV: the generation of a stochastic gravitational
wave background (SGWB) within the sensitivity of the Laser Interferometer Space Antenna (LISA) and,
simultaneously, primordial magnetic fields that would evolve through the Universe's history and could be compatible
with the lower bound from $\gamma$-ray telescopes on intergalactic magnetic fields (IGMF) at present time.
}
{The SGWB spectrum is evaluated adopting semi-analytical models,
accounting for both the contributions from sound waves and magnetohydrodynamic (MHD)
turbulence in the aftermath of the first-order phase transition.
Turbulence is assumed to arise only after an initial
period of sound waves,
and the magnetic field is assumed to be in equipartition with 
the turbulent kinetic energy.
Several paths are considered for the magnetic field evolution throughout the radiation-dominated era,
in order to predict the amplitude and correlation length scale of the resulting IGMF today. 
Comparing the SGWB level with the sensitivity of LISA  and the IGMF parameters with the sensitivity 
reach of the CTA \gr\ telescope, we identify a range of
first-order phase transition parameters 
providing observable signatures at both detectors.
}
{
 If even a small fraction of the  kinetic energy in sound waves is converted into MHD turbulence,
 a first-order phase transition 
 occurring at a temperature between $1$ and $10^6$ GeV
 can give rise to an observable SGWB signal in LISA and, at
 the same time, an IGMF compatible with the lower bound from the $\gamma$-ray telescope MAGIC,
 for all proposed evolutionary paths of the magnetic fields throughout  the radiation-dominated era
 (i.e., for both helical and non-helical magnetic fields).
 For the following fractions of the energy density converted into 
 turbulence, $\varepsilon_{\rm turb}=0.1$ and $1$, we
 provide the range of first-order phase transition parameters (strength $\alpha$, duration $\beta^{-1}$,
 bubble wall speed $v_w$, and temperature $T_*$), together with the corresponding range of magnetic
 field strength $B$ and correlation length $\lambda$, 
 which would lead to the 
 SGWB and IGMF being observable with LISA and MAGIC. 
 The resulting magnetic field strength at recombination can also correspond 
 to the one that has been proposed to induce baryon clumping,
 previously suggested as a possible way to ease the Hubble tension.
 In the limiting case $\varepsilon_{\rm turb} \ll 1$, the SGWB is only sourced by
 sound waves, but an IGMF is still generated
 with a magnetic
 field amplitude proportional to $\sqrt{\varepsilon_{\rm turb}}$.
 We find that for values as small as $\varepsilon_{\rm turb} \sim {\cal O}(10^{-13})$ or
${\cal O} (10^{-9})$, respectively, helical or non-helical
 primordial magnetic fields
 provide IGMF compatible with MAGIC's lower bound.
 }
{}

\maketitle

{
}

\section{Introduction}

Magnetic fields are ubiquitous at all scales in the Universe.
In particular, observations from TeV blazars suggest the existence of intergalactic magnetic fields (IGMF)
in the cosmic voids of the large scale structure (LSS) with a lower bound in their strength of
$B > 1.8 \times 10^{-17}$ gauss at scales $\lambda_B \gtrsim 0.2$ Mpc \citep{Neronov:2010gir,MAGIC:2022piy}.
Although the origin of these IGMF is still unclear, a seed of cosmological origin might be the most plausible 
explanation,
since astrophysical mechanisms (e.g., via Biermann battery) seem to lack the potential to generate IGMF in
cosmic voids with a large volume filling factor \citep{2011ApJ...727L...4D}.

The presence of a primordial magnetic field in the early Universe is, on the other hand,
constrained by the Big Bang nucleosynthesis (BBN) limit, first analysed in \citet{Shvartsman:1969mm},
as well as by the cosmic microwave background (CMB) experiments \citep{Planck:2015zrl,Galli:2021mxk}.
Several mechanisms have been proposed that would lead to the generation of magnetic fields from cosmological
phase transitions, especially if the phase transition is of the first order; see \citet{Durrer:2013pga} and \citet{Vachaspati:2020blt} for reviews and references therein.
According to the Standard Model (SM), both the electroweak and the quantum chromodynamics (QCD)
transitions have occurred as crossovers in the early Universe \citep{Kajantie:1995kf,Stephanov:2006zvm}.
However,  extensions of the SM can provide the required conditions for baryogenesis at the electroweak
scale and would produce, at the same time, a first-order phase transition; see \cite{Caprini:2019egz} for a review.

First-order phase transitions, as well as turbulent motion,
occurring around the electroweak scale, can be responsible for the production of a
stochastic gravitational wave background (SGWB); see, for example, \cite{Kamionkowski:1993fg},
detectable by the Laser Interferometer Space Antenna (LISA; \citealt{LISA:2017pwj,LISA:2024hlh}).
Due to the high conductivity of the primordial plasma in the early Universe,
a magnetic field would interact with it, inevitably leading to the development of magnetohydrodynamic (MHD)
turbulence \citep{Brandenburg:1996fc}.
The production of gravitational waves (GWs) from (M)HD turbulence has been extensively
studied (see \cite{Caprini:2018mtu} for a review and references therein), and
numerical simulations \citep{RoperPol:2019wvy} have provided the tools to compute
and validate an analytical template for the SGWB \citep{RoperPol:2022iel}.

Primordial magnetic fields generated
at a phase transition, either non-helical \citep{Vachaspati:1991nm} or helical
\citep{Vachaspati:2001nb},
have a characteristic length scale that
is limited by the size of the horizon, for example, $\tilde \lambda_{H} \sim\!10^{-9}$~Mpc
(redshifted to present time due to the expansion of the Universe) for a temperature scale of $T_*\sim100$ GeV.
The magnetic field follows a dynamical evolution
driven by
decaying turbulence from the time of its generation until 
the end of the radiation-dominated era.
Later, the characteristic turbulent power-law evolution of the magnetic
field strength and length scale 
became logarithmic functions of cosmic time, such that the field
evolution at large scales is virtually frozen \citep{Banerjee:2004df,Subramanian:2015lua}.

The details of the field evolution during the radiation-dominated era are not fully understood. 
If the turbulent decay of non-helical magnetic fields from the electroweak epoch
proceeds through the "selective decay" of short-range modes  \citep{Banerjee:2004df},
its present-day relic strength is expected to be below the lower bound from the \gr\ observations
when the redshifted size of the horizon is around $10^{-9}$ Mpc or smaller (for $T_* \gtrsim 100$ GeV)
and the comoving initial field strength is limited by the BBN bound $\tilde B \lesssim 10^{-6}$ G
\citep{Kahniashvili:2009qi}. 
In contrast, helicity conservation of decaying helical magnetic fields
leads to an inverse cascade that amplifies the magnetic field length scale and the field amplitude at large scales,
while the magnetic energy density decays \citep{Banerjee:2004df}.
Hence, helical fields are compatible with the IGMF inferred from the \gr\ observations 
for a larger set of initial conditions.
However, it has been argued from simulations that a similar large-scale field amplification,
although less effective than for fully helical fields,
can also occur for non-helical fields \citep{Brandenburg:2014mwa}.
This has also later been derived using theoretical arguments: 
\cite{Hosking:2022umv} propose that the conservation of local 
helicity fluctuations, characterised by the so-called Hosking 
integral \citep{Zhou:2022xhk}, can lead to an inverse cascade even when the net helicity is zero.
Therefore, fields produced around the electroweak scale with
any fractional helicity could,
in principle, provide a seed field compatible with the \gr\ lower bounds of the IGMF.

In addition, \cite{Jedamzik:2020krr} proposed that a primordial magnetic field present at recombination,
with sufficient amplitude, could reduce the sound horizon by inducing baryon clumping, which would affect
the estimate of the present-time Hubble constant from CMB data in a way that would relax the "Hubble tension."
\cite{Neronov:2021xua} and \cite{RoperPol:2022iel} showed that a non-helical field originating in the QCD
phase transition could satisfy the \gr\ lower bound of the IGMF, induce enough baryon clumping
\citep{Jedamzik:2020krr,Galli:2021mxk}, 
and generate the SGWB that has been detected by several pulsar timing array
collaborations
(\citealt{NANOGrav:2023gor,EPTA:2023fyk,Reardon:2023gzh,Xu:2023wog}; see, in particular, \citealt{EPTA:2023xxk}, where (M)HD turbulence was analysed as a possible source of the SGWB,
and \citealt{NANOGrav:2023hvm}, where  sound waves and bubble collisions were analysed instead).

In the present work, we provide a framework for studying a first-order phase transition with a
"multi-messenger" approach.
On the one hand, we analyze the detectability by LISA of a SGWB produced around the electroweak scale,
from both sound waves \citep{Hindmarsh:2013xza} and (M)HD turbulence.
On the other hand, we analyze the evolution down to recombination of a 
magnetic field also generated around the electroweak scale, compatible with the aforementioned SGWB,
and determine whether it can explain the blazars observations in the cosmic voids of the LSS,
the CMB and BBN constraints, as well as the possibility that it can alleviate the Hubble tension.
All the results and models presented can be reproduced using the {\sc CosmoGW} package
that is publicly available on GitHub \citep{CosmoGW_GH}.

In the following, the notation is such that all characteristic scales and time intervals are physical
and, therefore, time-dependent. They are understood to be redshifted when compared with quantities at
the present time, for example, with the GW frequency $f$ or with the conformal Hubble factor at the phase transition time,
${\cal H}_* \equiv (a_*/a_0)\, H_*$,
$H_\ast$ being the Hubble rate and $a_\ast/a_0$ the ratio of the scale factor
at the time of GW production with that at the present time.

\section{Models for the stochastic gravitational wave background}
\label{background_templates}

We study the potential detectability of an SGWB from a first-order phase transition around the electroweak scale,
incorporating updated results on the contributions to the SGWB
from sound waves \citep{Hindmarsh:2019phv,Jinno:2022mie} and (M)HD turbulence \citep{RoperPol:2022iel}.
Since it has been found that the electroweak phase transition bubbles hardly run away \citep{Bodeker:2017cim},
we ignore the contribution to the SGWB from broken-phase bubble collisions.

Following \citet{Caprini:2015zlo,Caprini:2019egz}, we characterise the phase transition
by four parameters. 
\begin{enumerate}
    \item $T_*$ is the phase transition temperature (interpreted as the percolation temperature for a weak to
    moderately strong thermal phase transition).
    \item $\beta$ is the rate of change in the nucleation rate of broken-phase bubbles.
    Its inverse value at the moment of the phase transition can be considered as a characteristic timescale of
    the phase transition, $t_\beta=\beta^{-1}\lesssim H_*^{-1}$,
    where $H_*$ is the expansion rate at the epoch
    of the phase transition.
    \item $\alpha$ characterises the strength of the phase transition: it  represents the vacuum energy density
    in the symmetric phase released in the broken-phase bubbles, in units of the radiation energy density\footnote{Note
    that this parameter has also been generalised beyond the bag equation of state using, for example, the
    trace anomaly of the energy momentum tensor or the difference among the energies in the two phases
    \citep{Caprini:2019egz}. 
    In practice, for the purpose of this work, all three definitions would be equivalent.}
    $\alpha =\rho_{\rm vac}/\rho_{\rm rad}^*$. \vspace{1mm}
    \item $v_w$ is the velocity of the broken-phase bubbles.
\end{enumerate}
Apart from these dynamical parameters, the inclusion of (M)HD turbulence in the calculation of the SGWB spectrum
adds another parameter:
\begin{enumerate}
\setcounter{enumi}{4}
    \item $\varepsilon_{\rm turb}$ is the fraction of the kinetic energy density induced in the primordial
    plasma by the expansion and collisions of bubbles that is converted from sound waves into vortical turbulent motion. 
\end{enumerate}
In the analysis of possible SGWB spectra from a cosmological first-order phase transition, we scan the four-dimensional
parameter space $(T_*,\beta,\alpha,v_w)$ and consider two possible levels of $\varepsilon_{\rm turb} = 0.1$ and 1.
The case $\varepsilon_{\rm turb} = 0$, where the SGWB is dominated by sound
waves, is analysed by the LISA Cosmology Working Group in \cite{Caprini:2024hue}.

The resulting SGWB spectrum is the result of the addition of two contributions: sound waves and (M)HD turbulent plasma motion.
However, in the limiting case $\varepsilon_{\rm turb} = 0$, the contribution
of (M)HD turbulence is absent.
The interplay between these two contributions is uncertain, and it is
an open, active topic of research \citep{Caprini:2024gyk,Correia:2025qif}.
For example, a mixed compressional-vortical mode
would appear in the GW spectrum when
both sound waves and turbulence are present
in the fluid perturbations, which has not yet been
studied in detail (see \cite{Correia:2025qif} for
a recent numerical study of this term).
In this work, we assume that sound waves (compressional motion) produce GWs from the
time of bubble collision 
until the development of nonlinearities at $\tau_{\rm nl}
\sim
\lambda_\ast/v_f$, where $v_f$ is the enthalpy-weighted
rms fluid velocity \citep{Caprini:2024hue}
and $\lambda_*$ is the characteristic length scale of the fluid motion (see Sec.~\ref{gw_sw}).
After this time, vortical (M)HD turbulence is produced due to the development of nonlinearities from the initial sound waves,
yielding an additional contribution to the SGWB, and the
remaining compressional motion is assumed to have dissipated.
Therefore, values of $\varepsilon_{\rm turb}$ smaller than unity would take into
account the decay of sound waves from their time of formation to the time of the development of
nonlinearities and the potential additional energy dissipation due to the development of turbulence
\citep{Caprini:2024gyk}.
We assume that the turbulent sourcing of GWs operates only in the decaying phase, which amounts
to assuming that the development of turbulence is instantaneous. 
Since the  details of the turbulence generation from the nonlinearities and shocks after the phase dominated by  sound waves
 are unknown, we remove this potential source of uncertainties in the 
determination of the resulting SGWB \citep{RoperPol:2019wvy,RoperPol:2021xnd,Brandenburg:2021tmp,RoperPol:2022iel}.
Furthermore, we omit the effect of reheated droplets
that can be produced when bubbles collide in strong
deflagrations, as found in \cite{Cutting:2019zws}.
This effect can significantly impact the GW amplitude, but
it is still uncertain how it depends
on the phase transition parameters.

Alternatively, both compressional and vortical motion (i.e., the development of turbulence) could  
occur directly together at
the time of bubble collisions \citep{Caprini:2024gyk,Correia:2025qif}.
In this scenario, the energy density of each contribution would directly come from the
energy budget available from the first-order phase transition
(for example, a fraction $\gamma$ of the total kinetic energy transferred to the fluid by the bubble expansion would go into
compressional or acoustic modes, and a fraction $1 - \gamma$ into
vortical modes).
The limiting case $\gamma = 0$, corresponding to only turbulence being produced at the phase transition,
was considered in \cite{RoperPol:2022iel}.
This case could model a scenario where the magnetic
field is produced by an alternative mechanism, instead of being amplified by the turbulent motion following
an initial development of sound waves, for example, when the magnetic field is
directly produced by bubble collisions \citep{Zhang:2019vsb} or driven by a scalar
axion-like field \citep{Miniati:2017kah}.
We compare the results of our analysis, in which we always account for
the initial compression phase characterised by sound waves, with the results of
\cite{RoperPol:2022iel} in Sec.~\ref{results_analysis}.

For the sound-wave component, we considered two templates.
The first is based on the original
sound shell model (SSM) of \citet{Hindmarsh:2016lnk} and \cite{Hindmarsh:2019phv}, including the correction
for short shock times \citep{Caprini:2019egz}.
In this model, the spectrum has two breaks, one at the frequency corresponding to the inverse mean size of
the broken-phase bubbles, $\lambda_*^{-1}$, and one at the scale corresponding to the inverse sound shell thickness,
$\delta \lambda_*^{-1} \geq \lambda_*^{-1}$. 
To ensure the correct hierarchy of scales proposed within the SSM, we fixed $v_w > 0.3$, since lower $v_w$
would yield $\delta \lambda_*^{-1} \leq \lambda_*^{-1}$; see, in comparison ~Eq.~(\ref{Deltaw}).
The spectral shape is a double broken power law, proportional to $f^9$ at small $f \lesssim \lambda_*^{-1}$,
proportional to $f$ at intermediate $\lambda_*^{-1} \lesssim f \lesssim \delta \lambda_*^{-1}$,
and proportional to $f^{-4}$ at large $f$.
Since we rely on analytical templates, in this work
we do not account for the fact that the SSM shape, in particular the intermediate slope, likely has
a more complicated dependence on $v_w$, as shown in
\citet{Gowling:2021gcy,RoperPol:2023dzg}.

\begin{figure}
    \centering
    \includegraphics[width=\columnwidth]{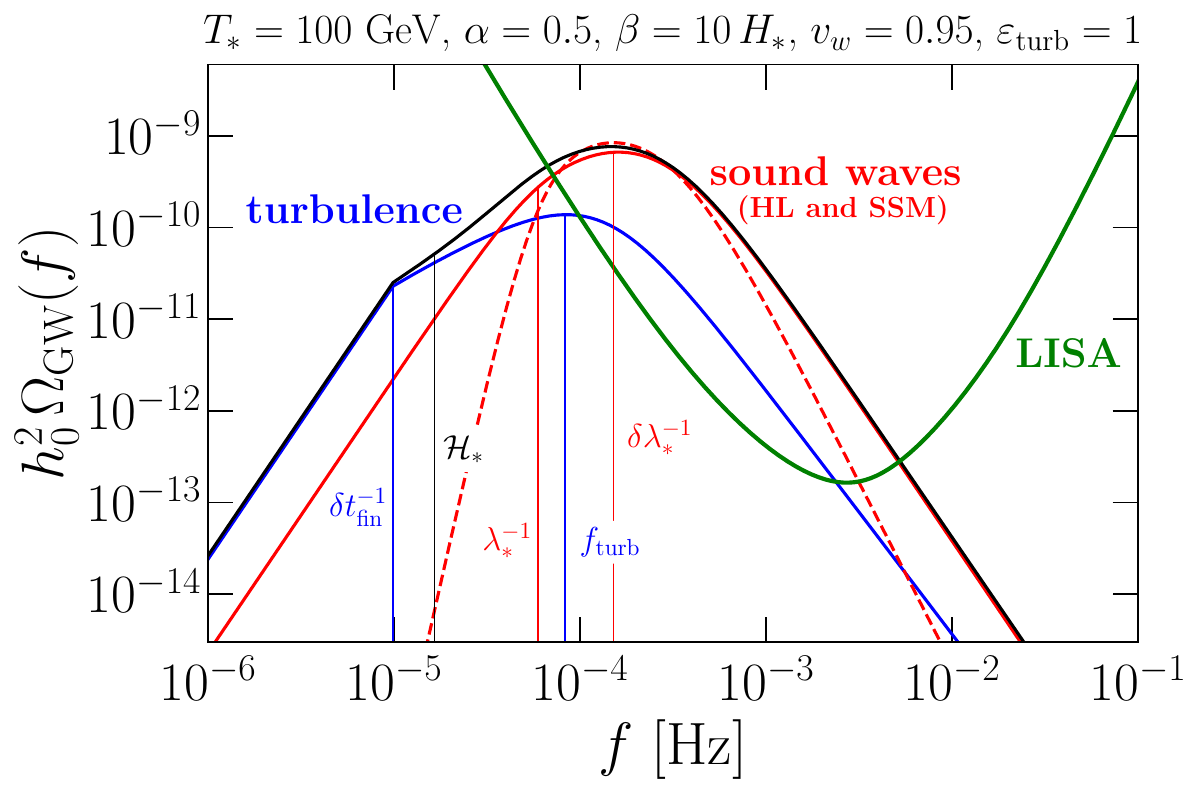}
    \caption{Spectra of the different components of the SGWB for $T_*=100\mbox{ GeV}$, $\alpha=0.5$,
    $\beta=10H_*$, $v_w = 0.95$, and $\varepsilon_{\rm turb}=1$, compared to LISA's power law sensitivity
    from \citet{Caprini:2019egz} with an S/N threshold of 10. 
    The SGWB spectra from sound waves are based on the SSM of \citet[dashed red]{Hindmarsh:2019phv} and on the
    fit from HL simulations of \cite{Jinno:2022mie}
    and \citet[solid red]{Caprini:2024gyk}.
    The spectrum of the turbulence is based on the model developed and numerically validated
    in \cite{RoperPol:2022iel}. 
    The vertical lines indicate the relevant frequencies of the turbulence template ($\delta t_{\rm fin}^{-1}$ and $f_{\rm turb}$) in blue, of the sound-wave template ($\lambda_*^{-1}$ and $\delta \lambda_*^{-1}$) in red, and the conformal
    Hubble frequency ${\cal H}_*$ at the time of GW generation in black.}
    \label{fig:example}
\end{figure}

After a revision of the sound shell model in \cite{RoperPol:2023dzg},
the $f^9$ part of the spectrum has been shown to appear at
$f \lesssim \lambda_\ast^{-1}$ only when the duration of the sound-wave
sourcing is long, $\tau_{\rm nl}/\lambda_\ast \sim 1/v_f \gg 1$.
The authors also show that this steep spectrum is followed by a linear
and then a cubic scaling at smaller frequencies.
The finding of a cubic scaling below $\lambda_\ast^{-1}$ is compatible with
numerical studies of phase transitions following the Higgsless (HL) approach
\citep{Jinno:2022mie,Caprini:2024gyk}, with the numerical study of sound
waves using the {\sc Pencil Code} \citep{PencilCode:2020eyn} by \cite{Sharma:2023mao},
and with the analytical analysis by \cite{Cai:2023guc}.
Hence, we considered a second template based on these updated results:
a spectrum proportional to $f^3$ is considered at small frequencies, with $f$ and
$f^{-3}$ scaling at intermediate and large $f$, respectively.
For simplicity, we omitted the steep spectrum below the peak, since
its prominence is dependent on the duration of the sound-wave regime and it is expected
to be mostly relevant for weak phase transitions, which are of less importance for
a potential detection with LISA.
The behavior at large frequencies is not in contradiction with that of \cite{Hindmarsh:2019phv} since their fit
asymptotically tends to $f^{-4}$ but their numerical results
show a $f^{-3}$ spectrum, their fit being accurate at small and
moderate $f$.
The differences between the two templates, which are described in
Sec.~\ref{gw_sw}, are illustrated in Fig.~\ref{fig:example}.

The SGWB of the turbulent component is described in Sec.~\ref{gw_mhd}. It is based on the model
presented and validated against simulations of MHD turbulence in
\citet{RoperPol:2022iel}.
This model has also been validated for purely kinetic turbulence in \cite{Auclair:2022jod}.
Although this model has been tested for non-helical fields,
the numerical simulations in \cite{RoperPol:2021xnd} indicate that the SGWB sourced by decaying turbulence
does not significantly depend on the fractional helicity.
Therefore, we applied this template to the case of
helical magnetic fields as well, in particular when we considered the time evolution of the field
(cf.,~Fig.~\ref{fig:B_limits}).

The turbulent spectrum also presents two breaks.
In this case, the first break corresponds to the inverse effective duration of the turbulence $\delta t_{\rm fin}^{-1}$,
and the second is $f_{\rm turb} \sim {\cal O} (\lambda_*^{-1})$.
The first break in the GW spectrum is from $f^{3}$ to $f$, and the second break (the peak of the spectrum) is
from $f$ to $f^{-8/3}$, where 8/3 is characteristic of Kolmogorov turbulence \citep{RoperPol:2019wvy}.
Note that this template applies under the assumption that turbulence is generated instantaneously. 
The resulting sound wave and turbulence templates are shown in Fig.~\ref{fig:example}
for a specific choice of the parameters describing the phase transition.
The templates described in this section, both for sound waves and for
(M)HD turbulence, have been included in the publicly available Python package
{\sc CosmoGW} \citep{CosmoGW_GH}, where a tutorial is also available to follow
the results presented in our work.

\subsection{Gravitational wave background from sound waves}
\label{gw_sw}

The template of the spectrum of GWs from sound waves relies on the updated analysis of \citet{Hindmarsh:2019phv},
within the framework of the SSM \citep{Hindmarsh:2016lnk}. 
Within this model, the GW power spectrum has two characteristic scales.
The first one is the mean size of the broken-phase bubbles at the moment of percolation
\begin{equation}
    \lambda_{*}=\frac{(8\pi)^{1/3}\max(v_w,\cs)}{\beta},
    \label{lambda_star}
\end{equation}
where the sound speed is taken to be $\cs = 1/\sqrt{3}$ during the radiation-dominated era.
We added the correction at low $v_w$, following \cite{Caprini:2019egz} and \cite{Ellis:2020awk}. 
The second scale is the sound shell thickness \citep{Hindmarsh:2019phv}
\begin{equation} \label{Deltaw}
    \delta\lambda_*=\Delta_w\lambda_*,\ \ \text{with \ }  \Delta_w=\frac{|v_w-\cs|}{v_w}.
\end{equation}
We adopted the following
analytic expression to approximate the SSM SGWB spectrum  \citep{Hindmarsh:2019phv,Hindmarsh:2020hop}
\begin{equation}
    \Omega_{\rm GW}^{\rm sw} (f) = 3\, {\cal B} (\Delta_w) 
    \frac{(K \lambda_* {\cal H}_*)^2}{\sqrt{K} + \lambda_* {\cal H_*}}
    F_{\rm GW}^0 \,
    S_{\rm sw} (s_1, s_2, \Delta_w),
    \label{OmGW_SW}
\end{equation}
where $S_{\rm sw}$ is the spectral shape, given in Eq.~(\ref{spectral_shape_SSM}) for the SSM and
in Eq.~(\ref{spectral_shape_HL}) for the fit based on HL simulations;
$s_1 = \lambda_* f$ and $s_2 = \delta \lambda_* f$ are the frequencies expressed in units of each of the relevant scales.
The GW efficiency is ${\cal B} = \tilde \Omega_{\rm gw}/
    \mu (\Delta_w),$ where
$\tilde \Omega_{\rm gw} \sim {\cal O}
(10^{-2})$ is the integrated value of the spectrum, based on numerical simulations \citep{Hindmarsh:2017gnf}.
The normalisation variable $\mu$ is defined to ensure that the integral of $S_{\rm sw}/\mu$ over $\ln f$ is unity.
It is a function of the sound shell thickness
\begin{equation}
    \mu (\Delta_w) = \int_{0}^\infty 
    S_{\rm sw} (f, \lambda_*, \Delta_w) \,
    \dd \ln f.
\end{equation}
$K$ denotes the amount of sound-wave kinetic energy density as a fraction of the total energy density
\begin{equation}
    K = \frac{\rho_{\rm K}}{\rho_{\rm vac} + \rho_{\rm rad}^*} =
    \frac{\kappa_v \alpha}{1 + \alpha},
    \label{K_def}
\end{equation}
where $\kappa_v \equiv \rho_{\rm K}/\rho_{\rm vac}^*$ is the fraction of the vacuum energy released that
is converted to the kinetic energy density of the plasma $\rho_{\rm K}$ \citep{Espinosa:2010hh}.
Therefore, in Eq.~(\ref{K_def}) we make the implicit assumption that the totality of the kinetic energy available
from the phase transition is in the form of sound waves. 
In the simulations of~\cite{Hindmarsh:2017gnf}, it has been demonstrated that this is a good assumption for small $\alpha$.
We used the results of \citet{Espinosa:2010hh} to calculate $\kappa_v$ as a function of $\alpha$ and $v_w$.

The spectral shape $S_{\rm sw}$ in the SSM is a function of the 
frequency $s_2$, expressed in units of the global peak frequency
$\delta \lambda_*^{-1}$,
i.e., the inverse sound shell thickness
\begin{equation}
    S_{\rm sw}^{\rm SSM} (s_2, \Delta_w)= s_2^9
    \left(\frac{\Delta_w^4+1}{\Delta_w^4+s_2^4}\right)^2
    \left(\frac{5}{5-m+ms_2^2}
    \right)^{5\over 2},
    \label{spectral_shape_SSM}
\end{equation}
where $m=(9\Delta_w^4+1)/(\Delta_w^4+1)$ is defined
so that $S_{\rm sw}^{\rm SSM} = 1$ at the peak $s_2 = 1$.

On the other hand, following \cite{Jinno:2022mie}, we also
considered a double broken power law with slopes $3$, $1$, and $-3$ at small, intermediate,
and large frequencies, respectively, again expressed in terms of the
frequencies $s_1$ and $s_2$
\begin{equation}
    S_{\rm sw}^{\rm HL} (s_1, s_2, \Delta_w) = 16 \, s_2^3 \frac{(1 + \Delta_w^{-3})^{2\over 3}}{(1 + s_1^3)^{{2\over 3}}
    \, (3 + s_2^2)^{2}}.
    \label{spectral_shape_HL}
\end{equation}
The spectral shape of this fit is based on the results of the HL simulations of \cite{Jinno:2022mie}
and \citet[see also \citealt{Stomberg:2025kxf} for a summary of their results]{Caprini:2024gyk},
while the frequency breaks and the amplitude are taken from the SSM.
The spectral shape is normalised so that $S_{\rm sw}^{\rm HL} = 1$ at the peak $s_2 = 1$.

Finally, the transfer function $F_{\rm GW}^0$ that
considers the redshift of the GW energy density is
\begin{equation}
    F_{\rm GW}^0 = \biggl(\frac{a_*}{a_0}\biggr)^4
    \biggl(\frac{H_*}{H_0}\biggr)^2
    \simeq 1.64\, h_0^{-2} \times 
    10^{-5} \,
    \biggl(\frac{100}{g_*}\biggr)^{1\over3},
\end{equation}
where $h_0$
is the current Hubble rate in units of 100 km/s/Mpc.

\subsection{Gravitational wave background from (M)HD turbulence}
\label{gw_mhd}

To model the production of GWs from (M)HD vortical turbulence, we followed the results presented in
\cite{RoperPol:2022iel}: we assumed that the onset of turbulence is instantaneous and that, afterward,
the GW sourcing stresses are constant in time  during the GW production, i.e., for a time that we denoted
$\delta t_{\rm fin}$  (more details of this quantity are given below).
This assumption gives a good approximation to the SGWB signal evaluated by numerical simulations in which the
turbulence is inserted in the initial conditions, as demonstrated in \cite{RoperPol:2022iel} for MHD turbulence and
in \cite{Auclair:2022jod} for purely vortical kinetic turbulence.
The SGWB spectrum reads
\begin{equation}
    \Omega_{\rm GW}^{\rm turb} (f) = 3 \, {\cal A} \, \Omega_{*}^2  \, (\lambda_* {\cal H}_*)^2
    F_{\rm GW}^0 \,
    S_{\rm turb} (s_1, s_3).
    \label{eq:OmGWturb}
\end{equation}

The amplitude of the SGWB spectrum scales with the square of the fraction of the total turbulent energy density
$\Omega_*$, given by the sum of the vortical velocity component 
and the magnetic field. 
We assumed $\Omega_*$ to be a fraction $\varepsilon_{\rm turb}$ of $K$,
\begin{equation}
    \Omega_* =  \Omega_{B*} + \Omega_{v*}^\perp = \frac{\rho_{B*} + \rho_{v*}^\perp}{\rho_{\rm vac} + \rho_{\rm rad}^*} =
    \varepsilon_{\rm turb} K,
    \label{eq:epturb}
\end{equation}
where $\rho_{B*}$ is the magnetic energy density, $\rho_{v*}^\perp$ is the vortical component of the kinetic energy
density, and $\Omega_{B*}$ and $\Omega_{v*}^\perp$ are their ratios to the total energy density.
Note that $\Omega_{v*}^\perp$ refers to the kinetic vortical motion while $K$, defined in Eq.~(\ref{K_def}), refers to the kinetic compressional motion in the form of sound waves.

Following \cite{Banerjee:2004df}, we assumed that the fields are in equipartition,
so that $\Omega_{v*}^\perp \simeq \Omega_{B*} \simeq {1\over2} \Omega_*$.
We defined the characteristic velocity $u_*$ as the maximum of the enthalpy-weighted
rms fluid velocity $v_f$ and the Alfv\'en speed $v_{\rm A}$.
For a radiation-dominated Universe with pressure $p = \cs^2 \rho_{\rm rad} = {1 \over 3} \rho_{\rm rad}$,
the subrelativistic Alfv\'en speed
$v_{\rm A}$ is \citep{Brandenburg:1996fc,RoperPol:2025lgc}
\begin{equation}
    v_{\rm A}^2 = \frac{\langle B^2\rangle}{\langle \rho_{\rm rad} + p\rangle} =
    \frac{2 \, \Omega_{B}}{(1 + c_s^2)} = {3 \over 2} \Omega_B,
    \label{Alfven_fn}
\end{equation}
and $v_f$, considering only the vortical motion $v_\perp$, is
\begin{equation}
    v_f^2 =
   \frac{\langle{w \, v_\perp^2\rangle}}{\langle{w \rangle}}
   =
   \Omega_{v}^\perp,
 \label{urms}
\end{equation}
where $w = \rho_{\rm rad} + p$ is the enthalpy of the fluid.
From Eqs.~(\ref{Alfven_fn}) and (\ref{urms}), we find that the characteristic
velocity $u_*$, assuming equipartition, is
\begin{equation}
    u_* =
    \sqrt{\max \Bigl(\Omega_{v*}^\perp, {3 \over 2} \Omega_{B*} \Bigr)}
    \simeq \sqrt{{3\over4}\Omega_*}.
\end{equation}

The parameter $\delta t_{\rm fin}$ represents the effective turbulence duration, and we express it in terms of
a multiple ${\cal N}$ of the eddy turnover time $\delta t_{\rm fin} = {\cal N} \lambda_*/u_*$. 
According to the numerical simulations in  \cite{RoperPol:2022iel}, one can set ${\cal N} \simeq 2$. 
However, note that the value of ${\cal N}$ is a subject of ongoing study. 
The effective turbulence duration determines the position of the first break in the SGWB spectrum,
occurring at $f\sim \delta t_{\rm fin}^{-1}$.

The SGWB spectral shape in Eq.~\eqref{eq:OmGWturb} is given by $S_{\rm turb} (s_1, s_3)$,
where $s_3 = \delta t_{\rm fin} f$ denotes the frequency in units of the inverse effective duration of the turbulence.
We recall that $s_1 = \lambda_* f$: as usual, we assume that the initial correlation length of the turbulence
corresponds to the mean size of the broken-phase bubbles at the onset of their collisions,
$\lambda_*$, given in Eq.~(\ref{lambda_star}).
This determines the position of the global peak of the spectrum, located at $s_{\rm turb} \equiv \lambda_* f_{\rm turb} \gtrsim 1$.
The exact value of $s_{\rm turb}$ depends on the spectral shape of the anisotropic stresses $p_\Pi (s_1)$; see Eqs.~(\ref{p_pi_fit})--(\ref{p_pi_max}).

The spectral shape $S_{\rm turb}$ is determined both by the anisotropic stress power spectrum itself,
and by how the latter sources the GWs in time. 
Assuming that the sourcing process is constant in time over the
duration $\delta t_{\rm fin}$ provides the following SGWB spectral shape
\begin{equation}
    S_{\rm turb} (s_1, s_3) =
    \frac{4 \pi^2 \, s_1^3 \, {\cal T}_{\rm GW}
    (s_1, s_3)}
    {(\lambda_* {\cal H}_*)^{2} } \,
    \frac{p_\Pi (s_1)}{s_{\rm turb}\, p_\Pi (s_{\rm turb})},
    \label{S_turb}
\end{equation}
with ${\cal T}_{\rm GW}$ being \citep{RoperPol:2022iel}
\begin{align}
    {\cal T}_{\rm GW} (s_1, s_3) =
    \left\{\begin{array}{ll}
        \ln^2 \bigl[1 + {\cal H}_*
        \delta t_{\rm fin}/(2 \pi) \bigr], & \hspace{-1mm}
        \text{if \ } s_3 < 1, \\
        \ln^2 \bigl[1 + \lambda_* {\cal H}_*/(2\pi s_1)\bigr], &  \hspace{-1mm}
        \text{if \ } s_3 \geq 1.
    \end{array} \right.
\end{align}
$S_{\rm turb}$ is normalised so that its maximum is approximately
one.
Its exact value at the peak $f_{\rm turb}$, where $s_3=\delta 
t_{\rm fin}f_{\rm turb} > 1$, is
\begin{align}
    S_{\rm turb} (s_{\rm turb}) = 4 \pi^2 \biggl(\frac{s_{\rm turb}}{\lambda_* {\cal H}_*}\biggr)^2
\ln^2 \biggl[1 + \frac{\lambda_* {\cal H}_*}{2 \pi s_{\rm turb}}\biggr],
\end{align}
which asymptotically tends to 1 for $\lambda_* {\cal H}_* \ll 2 \pi s_{\rm turb}$.
Since $s_{\rm turb} \gtrsim 1$ and $\lambda_* {\cal H}_* \leq 1$, one can compute the value of
$S_{\rm turb}(s_{\rm turb})$ at 
$\lambda_* {\cal H}_* = 1$, to get $S_{\rm turb}(s_{\rm turb})\gtrsim S_{\rm turb}(1) \simeq 0.86$.
As $\lambda_* {\cal H}_*$ decreases, $S_{\rm turb}(1)$ quickly tends to 1, with a relative error below
1.6\% when $\lambda_* {\cal H}_* < 0.1$.
Hence, the assumption that $S_{\rm turb}$ is one at the frequency peak is accurate for most relevant cases.

The power spectrum of the anisotropic stresses, $p_\Pi(s_1)$,
depends on the power spectrum of the kinetic turbulence and the magnetic field.
Here we assume that they follow a von K\'arman spectral shape\footnote{The von K\'arman spectrum can be expressed using, for example, Eq.~(6) in \cite{RoperPol:2022iel}
with $\alpha = 6/17$.} \citep{doi:10.1073/pnas.34.11.530}, characterised by
a Batchelor spectrum $f^4$ at low $f$, expected from causality \citep{Durrer:2003ja}, and a turbulent Kolmogorov $f^{-5/3}$ spectrum at large $f$.
Within this assumption, the normalised (i.e., $p_\Pi (0)=1$) anisotropic stress power spectrum
can be approximated as
\begin{equation}
    p_\Pi (s_1) \simeq \biggl[1 + \biggl(\frac{s_1}
    {s_\Pi}
    \biggr)^{\alpha_\Pi} \biggr]^{-{11\over{3\alpha_\Pi}}},
    \label{p_pi_fit}
\end{equation}
where $\alpha_\Pi \simeq 2.15$ and $s_\Pi = 2.2$.
Compared to the numerical computations in \cite{RoperPol:2022iel} and \cite{Auclair:2022jod},
the maximum relative error of the analytical fit of Eq.~(\ref{p_pi_fit})
is less than 6\% for all frequencies $s_1 \leq 10$.
In larger $s_1$, the relative error of the fit increases.
However, since $p_\Pi \sim s_1^{-11/3}$, at higher frequencies $p_\Pi(s_1 > 10)$ has decayed below ${\cal O} (10^{-3})$.

Having normalised both $S_{\rm turb}$ and $p_\Pi$, the
GW efficiency from (M)HD turbulence, ${\cal A}$ in Eq.~\eqref{eq:OmGWturb},
becomes \citep{RoperPol:2022iel}
\begin{align}
    {\cal A} = &\, \frac{112}{15} \sqrt{\frac{3}{5 \pi^5}}
    \frac{\Gamma\bigl[{13\over6}\bigr] \Gamma^2
    \bigl[{17\over6}\bigr]}{\Gamma\bigl[{17\over3}\bigr]
    \Gamma^2 \bigl[{1\over3}\bigr]}
    s_{\rm turb} \,p_\Pi (s_{\rm turb}) \nonumber \\
    \simeq &\,
    2 \times 10^{-3} s_{\rm turb} \,p_\Pi (s_{\rm turb}),
    \label{AA_turb}
\end{align}
where $s_{\rm turb}$, the peak frequency of the GW spectrum, is located where $s_1\, p_\Pi (s_1)$ is maximal
\begin{subequations}
\begin{align}
    &\,  s_{\rm turb} = \lambda_* f_{\rm turb} = s_\Pi \biggl({3 \over 8}\biggr)^{1\over\alpha_\Pi}
    \simeq 1.4, \label{peak_freq} \\
    &\, p_\Pi (s_{\rm turb}) =  \biggl(\frac{11}{8}
    \biggr)^{-{11\over3\alpha_\Pi}} \simeq 0.6.
\end{align}
\label{p_pi_max}
\end{subequations}
Substituting these values into Eq.~(\ref{AA_turb}),
the GW efficiency becomes
\begin{equation}
    {\cal A} \simeq 1.75 \times 10^{-3}.
\end{equation}
Note that \cite{RoperPol:2022iel} uses a different
notation, where $s_{\rm turb} \equiv k_{\rm GW}/k_*$, and
our parameter ${\cal A}$ corresponds to
$s_{\rm turb} \,p_\Pi (s_{\rm turb}) \,
    {\cal C}/(8 \pi^2 {{\cal A}}^2)$ in their 
Eq.~(24).

\section{The SGWB and the magnetic field, relics of a first-order phase transition}
\label{results_analysis}

In this section, we investigate which regions in the parameter space of a first-order phase transition would lead to a 
SGWB detectable with LISA. 
We also compute the strength and correlation length scales of the corresponding magnetic field, which would result from the amplification of a seed field by the MHD turbulence in the aftermath of the phase transition. 
We are ultimately interested in assessing 
whether a first-order phase transition could produce a SGWB detectable with LISA and, at the same time, an IGMF consistent with existing observational bounds. 
We therefore perform a systematic scan over the first-order phase transition parameter space.
The SGWB is considered ``detectable'' by LISA if its spectrum exceeds the power law sensitivity (PLS) of LISA with a S/N threshold of S/N$=10$
% \\\LEt{A & A use S/N for signal to noise ratio; please amend here.***}
\citep{Caprini:2019egz} at any frequency $10^{-5}$~Hz $<f<0.1$~Hz. An illustrative example of the SGWB spectrum from turbulence and sound waves with LISA's sensitivity for the choice of parameters  $T_*=100\mbox{ GeV}$, $\alpha=0.5$, $\beta=10H_*$, $v_w = 0.95$, and $\varepsilon_{\rm turb}=1$ is shown in Fig.~\ref{fig:example}.
In this work, we focus on multi-messenger probes of a first-order
phase transition and primordial magnetic fields, and we provide tools
for the analysis that have been added to the {\sc CosmoGW} code \citep{CosmoGW_GH}.
For a more elaborate parameter reconstruction of the SGWB from first-order
phase transitions with LISA, we refer to \cite{Caprini:2024hue}.

\begin{figure*}
    \centering
    \includegraphics[width=.8\columnwidth]{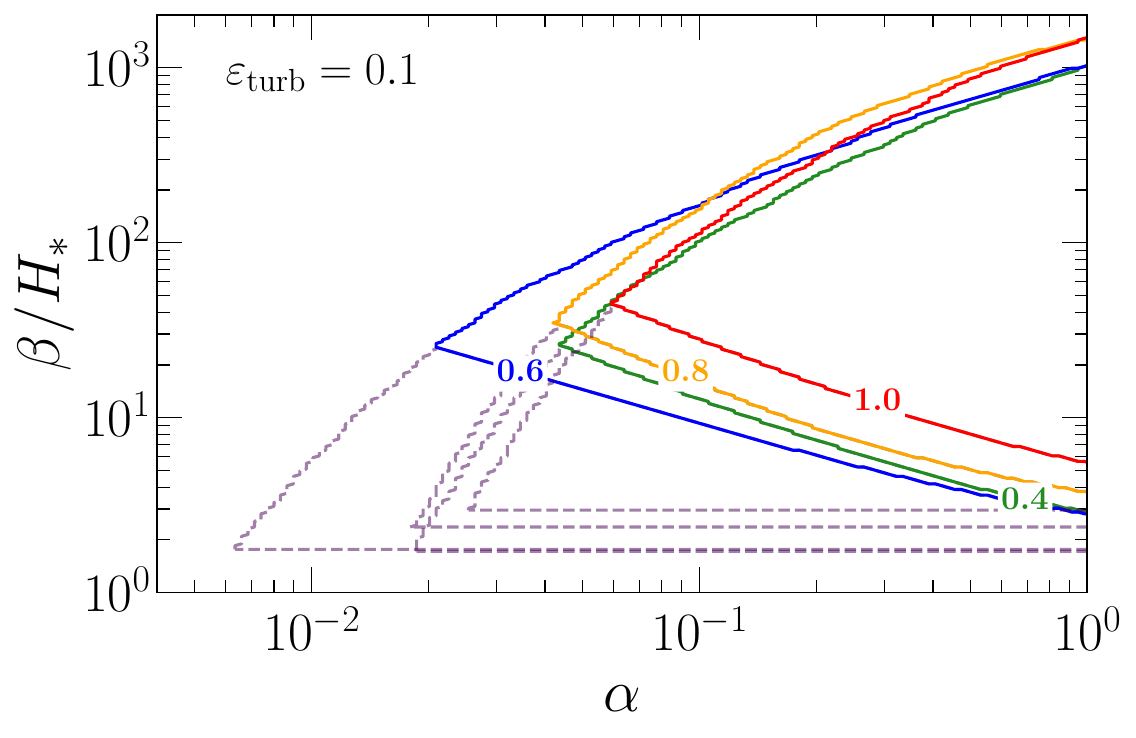}
    \includegraphics[width=.8\columnwidth]{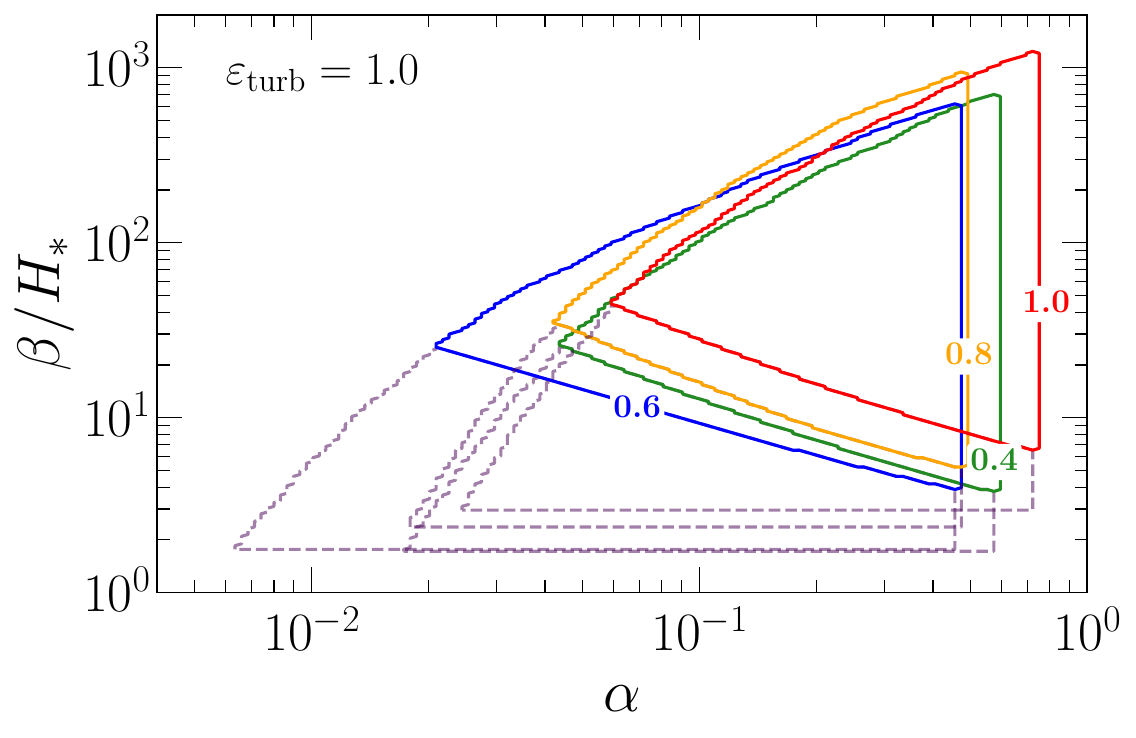}
    \includegraphics[width=.8\columnwidth]{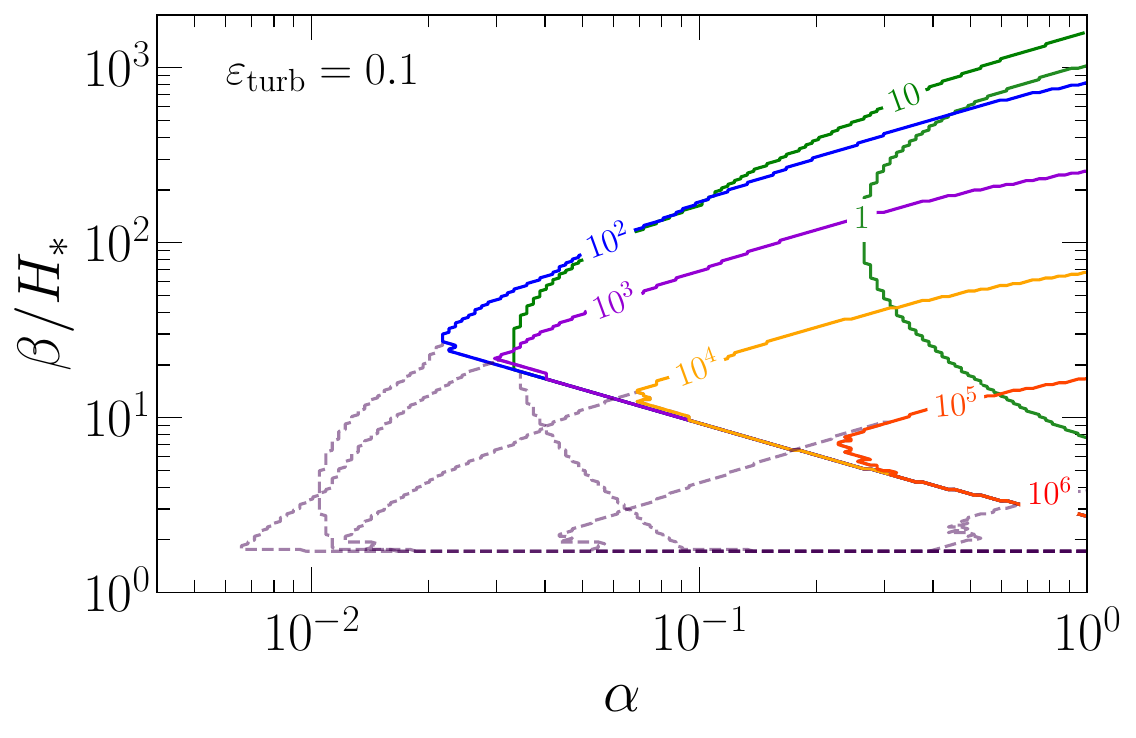}
    \includegraphics[width=.8\columnwidth]{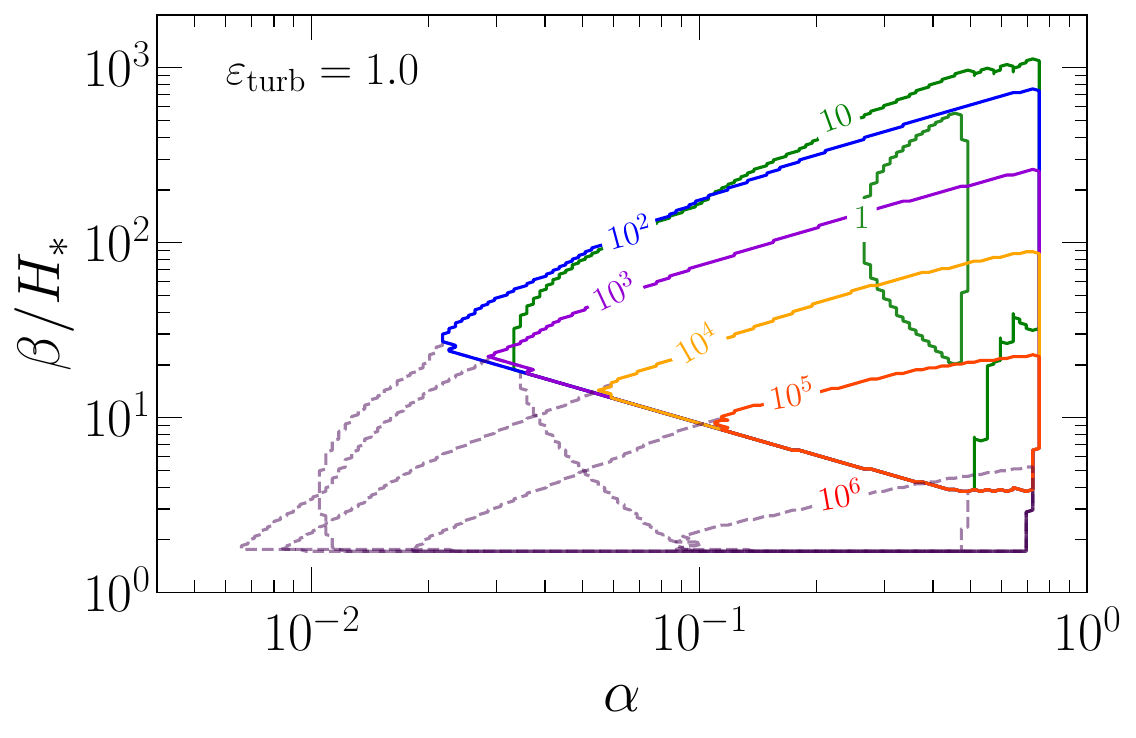}
    \caption{Contours of the parameters $\alpha$ and $\beta/H_*$
    that lead to a signal above LISA's power law sensitivity with a S/N threshold of 10 (see Fig.~\ref{fig:example}).
    The top panels take a fixed value of the wall velocity $v_w = \{0.4, 0.6, 0.8, 1\}$
    (indicated in different colors) and scan over temperatures to construct the
    contours, while the lower panels fix the temperature in GeV $T_\ast = \{10, 10^2,
    10^3, 10^4, 10^5, 10^6\}$ and scan over wall velocities to construct the contours.
    Left and right panels correspond to $\varepsilon_{\rm turb} = 0.1$ and 1, respectively.
    The results shown correspond to the sound-wave template based on the fit from the Higgsless simulations
    (cf.,~Sec.~\ref{gw_sw}), as this choice does not affect the magnetic
    field parameters in Fig.~\ref{fig:GW_B_signatures}.
    Colored contours correspond to regions of the parameter space where $H_* \tau_{\rm nl} \lesssim 1$, while their continuation by dashed contours corresponds to regions where $H_* \tau_{\rm nl} \gtrsim 1$.}
\label{fig:alpha_beta_GW_signatures}
\end{figure*}

We performed the analysis
for two values of $\varepsilon_{\rm turb}$, the parameter characterising
the relative amount of kinetic energy in (M)HD turbulence;
see, for comparison, ~Eq.~\eqref{eq:epturb}: $\varepsilon_{\rm turb}=0.1$ and 1.
The ranges of the phase transition parameters are defined as follows. 
The broken-phase bubble wall speed is in the range $0.3\leq v_w \leq 1$, where the
lower limit ensures that $\lambda_*>\delta\lambda_*$, and the hierarchy of breaks in the SSM SGWB spectral shape follows, as described in Sec.~\ref{background_templates}.
The upper limits on $\alpha$ are computed such that the
ratio of magnetic to total energy density, $\Omega_{B*}$, does not
exceed 10\%, as imposed by BBN \citep{Shvartsman:1969mm,Kahniashvili:2009qi,Kawasaki:2012va}.\footnote{Relaxing the BBN limit has been recently proposed in \cite{Kahniashvili:2021gym} by considering the turbulence decay from the magnetic field generation scale to BBN.
However, we still restrict $\Omega_{B*}$ in our analysis to avoid relativistic Alfv\'en and velocity speeds \citep{RoperPol:2025lgc}, a regime in which the SGWB templates have not been validated.}
The lower limits on $\beta/H_*$ are set so that the resulting length scale is causal $\lambda_* {\cal H}_* \lesssim 1$.
Furthermore, $T_*$, $\alpha$, and $\beta/H_*$  satisfy $1$~GeV~$\lesssim T_* \lesssim 3\times 10^6$~GeV, $\alpha \gtrsim 5 \times 10^{-3}$, and $\beta/H_* \lesssim 2 \times 10^3$,
as
inferred by the results of the analysis of the detectability of the SGWB signal.

For each point in the 4D parameter space, we reconstructed the SGWB signal and compared it with the sensitivity of LISA.
Figure~\ref{fig:alpha_beta_GW_signatures} shows the detectable regions in the  $(\alpha,\beta/H_*)$ plane in two ways: i) for different wall velocities $v_w$, by combining the results for all temperatures $T_*$ in the detectable range (upper panels), and ii) for different temperatures (in GeV), by combining the results for all $v_w$ (bottom panels).
The results shown in Fig.~\ref{fig:alpha_beta_GW_signatures} are derived using the HL fit for the sound-wave template of Eq.~\eqref{spectral_shape_HL}, since using the SSM template, given in Eq.~\eqref{spectral_shape_SSM}, does not lead to any appreciable
difference when we compute the resulting magnetic field parameters.

It is usually argued that turbulence is not produced if the timescale to develop nonlinearities from sound waves is larger than one Hubble time $H_* \tau_{\rm nl}  \sim H_* \lambda_*/v_f \gtrsim 1$ (cf., ~\cite{Caprini:2019egz}).
According to this criterion,
we show in Fig.~\ref{fig:alpha_beta_GW_signatures} the contours of the regions where turbulence is expected to develop by colored lines, while the dashed lines correspond to the contours of the regions in the parameter space that would be, in principle, excluded by this condition.

Plasma motion induced by the first-order phase transition can excite (M)HD turbulence, which could amplify any preexisting seed magnetic fields, for example,  generated by charge separation at the bubble walls \citep{Sigl:1996dm}.
Turbulence is expected to drive the magnetic field energy density to equipartition with the plasma kinetic energy density \citep{Banerjee:2004df,Durrer:2013pga}.
The equipartition argument suggests that the initial magnetic field configuration at the end of the phase transition is
$B_* \simeq \sqrt{2 \, \Omega_{B*} \, \rho_{\rm rad}^*} \simeq \sqrt{\Omega_* \rho_{\rm rad}^*}$ 
[see Eq.~\eqref{Alfven_fn}], where we assume that the total energy density is $\rho_{\rm vac} + \rho_{\rm rad}^* \simeq \rho_{\rm rad}^*$
for weak and moderately strong phase transitions. 
After fixing the value of $\varepsilon_{\rm turb}$, using Eqs.~\eqref{K_def} and \eqref{eq:epturb} we can relate the magnetic field strength $B_*$ at the phase transition time, assuming that both the turbulence generation and the establishment of equipartition are instantaneous, to the fraction of vacuum energy that is converted to kinetic energy $\kappa_v$, and ultimately to $\alpha$ using the prescription of \cite{Espinosa:2010hh}. 
Furthermore, the initial magnetic field correlation length would  correspond to the characteristic fluid scale $\lambda_*$, which is related to $v_w$, $c_s=1/\sqrt{3}$, and $\beta/H_*$ via Eq.~\eqref{lambda_star}.

\begin{figure}
    \centering
    \includegraphics[width=.9\columnwidth]{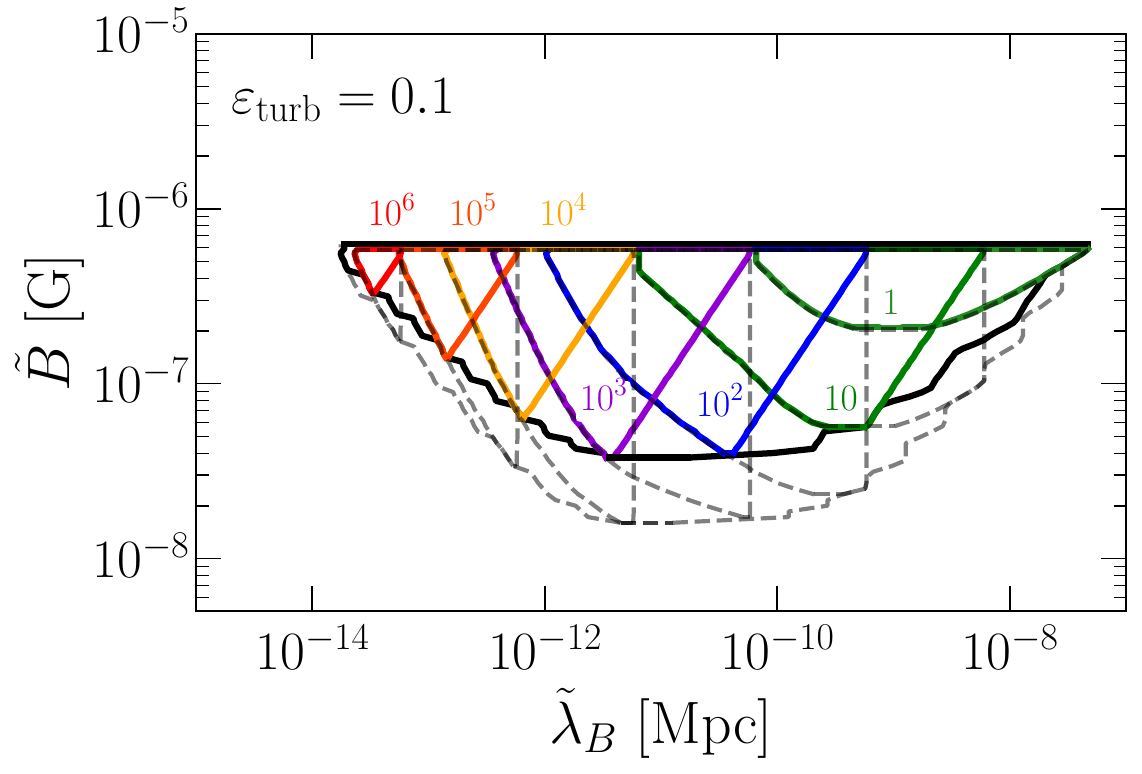}
    \includegraphics[width=.9\columnwidth]{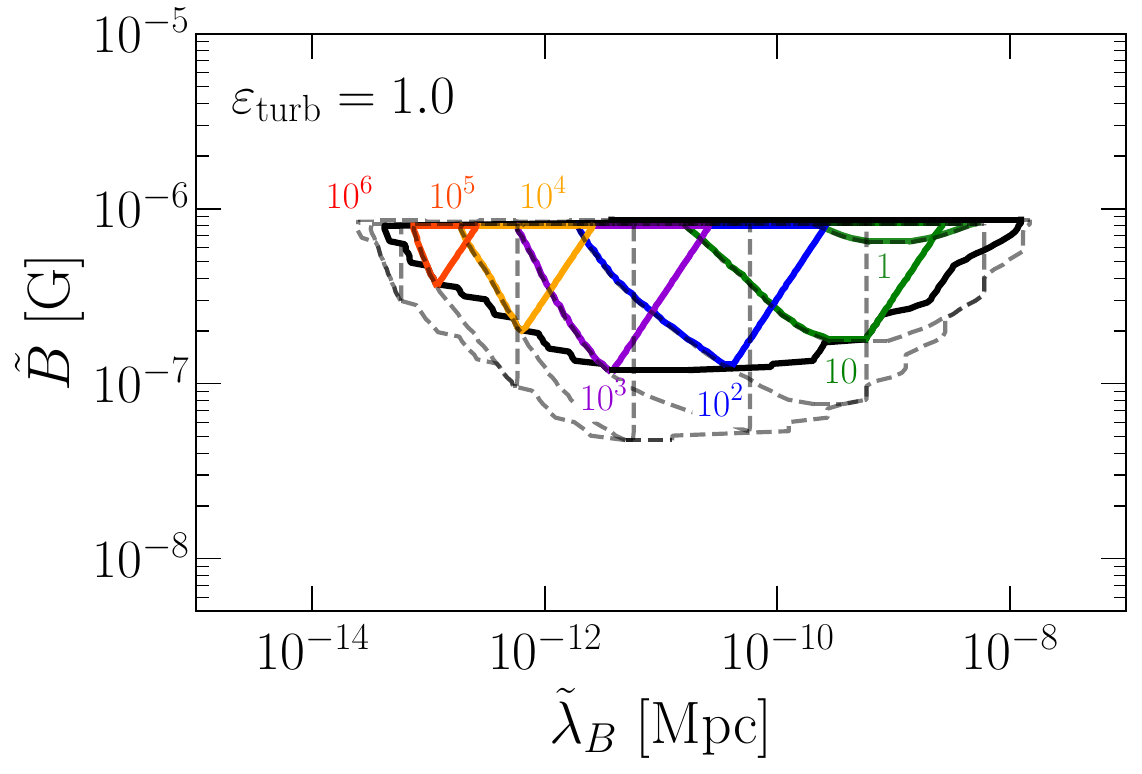}
    \caption{Range of the comoving magnetic field strength $\tilde B$ and correlation length $\tilde \lambda_B$ corresponding to the contours in the lower panels of Fig.~\ref{fig:alpha_beta_GW_signatures},
    for $\varepsilon_{\rm turb} = 0.1$ (top panel) and $1$ (bottom panel).
    The colored numbers indicate the value of the temperature $T_\ast$
    in GeV.
    For values of $\varepsilon_{\rm turb}$ smaller than 0.1, the range of observable 
    phase transition parameters is unaffected, and the contour of magnetic field
    parameters just shifts to smaller amplitudes proportional to $\sqrt{\varepsilon_{\rm turb}}.$
    }
    \label{fig:GW_B_signatures}
\end{figure}

Therefore, for each combination of the phase transition parameters analysed above, we can infer the magnetic field parameters $B_*$ and $\lambda_*$. 
As explained in Sec.~\ref{sec:evolB}, it is meaningful to express $B_*$ and $\lambda_*$ in terms of comoving quantities, $\tilde B$ and $\tilde \lambda_B$,
as these are the initial conditions for the magnetic field evolution, i.e., the starting points in the magnetic field evolutionary paths.  
In Fig.~\ref{fig:GW_B_signatures} we therefore show the ranges of the magnetic field parameters $\tilde B$ and $\tilde \lambda_B$  that correspond to the ranges of the phase transition parameters given in the bottom panels of Fig.~\ref{fig:alpha_beta_GW_signatures}, i.e., for different values of $T_*$.
Since $K < 1$ and due to equipartition, the
values of $\Omega_{B*}$ are limited to $0.5 \, \varepsilon_{\rm turb}$. 
In addition, we have limited its value to 10\% of the total energy density because of the BBN bound and the condition of non-relativistic turbulence. 
Therefore, in all the analyses presented $\Omega_{B*} \leq \min(0.1, 0.5 \, \varepsilon_{\rm turb})$.

It appears from Fig.~\ref{fig:GW_B_signatures} that a phase transition with parameters providing a SGWB accessible to LISA, and $\varepsilon_{\rm turb} = 1$ (upper panel), can lead to fairly strong magnetic fields with strengths up to $\tilde B\sim 10^{-6}$~G, corresponding to the limit $\Omega_{B*} \sim 0.1$.
For $\varepsilon_{\rm turb} = 0.1$ (bottom panel), the upper limit $\Omega_{B*} \sim 0.05$ yields instead a maximal field strength of $\tilde B \sim 6 \times 10^{-7}$~G.
The initial correlation length of the field can vary over a wide range, between $10^{-14}$ and $10^{-7}$~Mpc, depending on the temperature of the phase transition $T_*$, its duration $\beta^{-1}$, and  the velocity of the bubble wall $v_w$.
The region of parameter space where $H_* \tau_{\rm nl} \gtrsim 1$, 
indicated in Fig.~\ref{fig:alpha_beta_GW_signatures} by the dashed contours at small $\alpha$ and $\beta/H_*$, corresponds in  Fig.~\ref{fig:GW_B_signatures} to the region where $\tilde B$ is small and/or $\tilde \lambda_B$ is large.
As in Fig.~\ref{fig:alpha_beta_GW_signatures}, the contours in Fig.~\ref{fig:GW_B_signatures} correspond to the sound-wave template based on the fit from the Higgsless simulations (cf., ~Sec.~\ref{gw_sw}).
Using the SSM template would not lead to appreciable differences in the magnetic field 
parameters.

\section{The magnetic field evolution and the IGMF today}
\label{sec:evolB}

\begin{figure*}
    \centering
    \includegraphics[width=1.6\columnwidth]{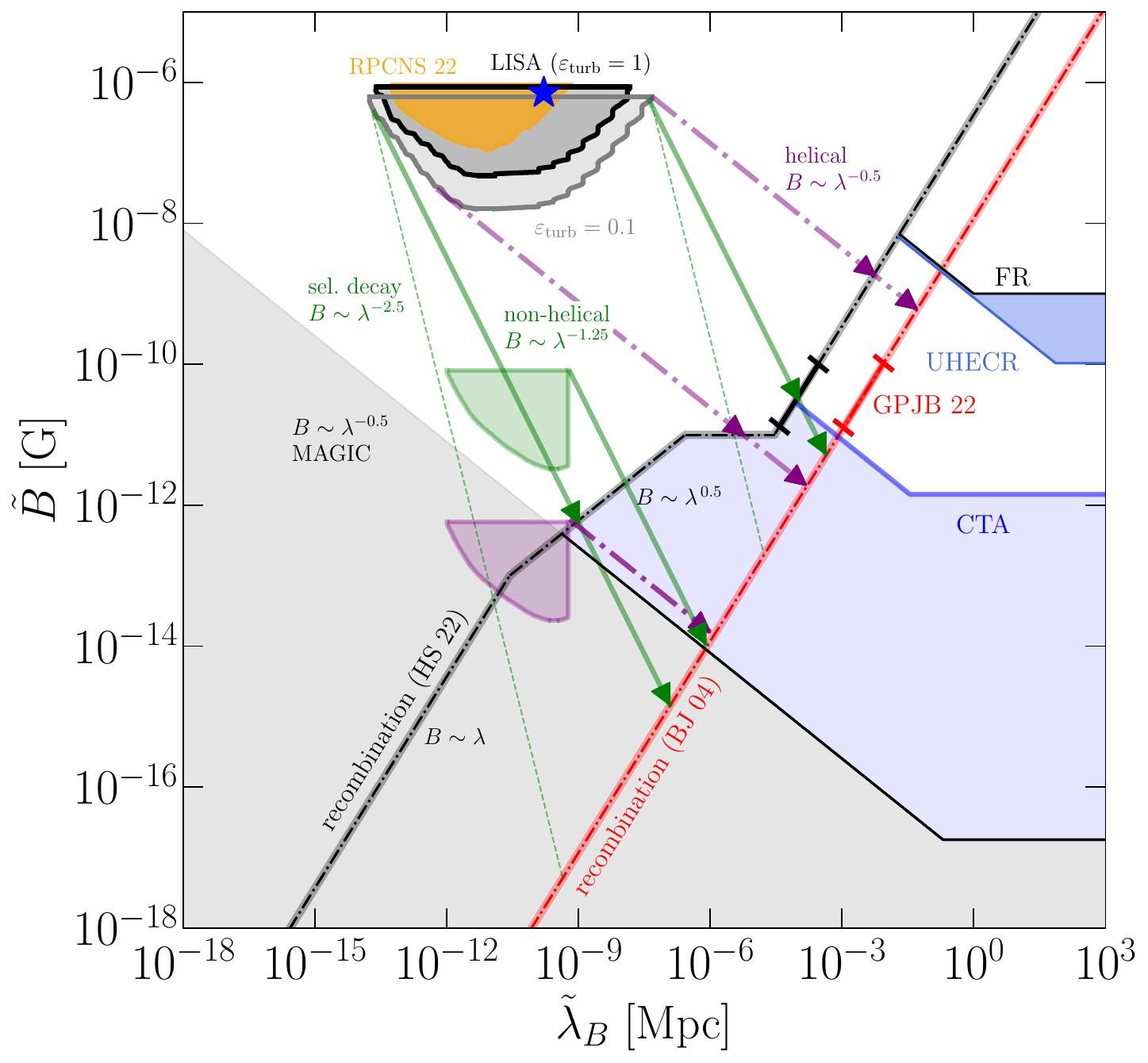}
    \caption{Expected initial and final comoving strength and correlation length of  a cosmological magnetic field generated at a first-order phase transition with parameters leading to an SGWB detectable by LISA.
    The black and grey contours in the upper left corner of the figure show the parameter space of initial conditions if one assumes that, respectively, the totality ($\varepsilon_{\rm turb}=1$) and 10\% ($\varepsilon_{\rm turb}=0.1$) of the sound-wave kinetic energy is eventually
    transferred to MHD turbulence
    (i.e., magnetic and vortical kinetic energy)
    at the time of nonlinearities development.
    The orange-shaded region is the one found in \citet[RPCNS 22]{RoperPol:2022iel},   assuming that the SGWB is sourced exclusively by MHD turbulence.
    The blue star corresponds to the phase transition parameters for which the SGWB spectrum is shown in Fig.~\ref{fig:example}, considered for illustrative purposes.
    The inclined arrows show the envelopes of the evolutionary tracks. The dash-dotted purple arrows apply to helical fields, $B \sim \lambda^{-0.5}$, as given in \cite{Banerjee:2004df}; the  solid green arrows apply to non-helical fields, $B \sim \lambda^{-1.25}$, as given in  \cite{Hosking:2022umv}; 
    the dashed green arrows correspond to the hypothesis of selective decay of short-range modes for non-helical fields, $B \sim \lambda^{-2.5}$ \citep{Banerjee:2004df}. 
    The dash-dotted black and red lines show the possible endpoints of the magnetic field evolution, corresponding to the IGMF  present in the voids today, assuming respectively that the reconnection timescale dominates the magnetic field dynamics \citep{Hosking:2022umv} (HS 22) and that the Alfv\'enic timescale dominates the magnetic field dynamics \citep{Banerjee:2004df} (BJ 04). 
     The gray-shaded region at the bottom left of the plot is excluded by  
    the lower bound on the IGMF established by the MAGIC \gr\ observatory \citep{MAGIC:2022piy}. 
    The thin dark blue and black lines show the upper limit on the IGMF from, respectively,  ultra-high-energy cosmic rays \citep{Neronov:2021xua} and Faraday rotation measures \citep{Pshirkov:2015tua}.
    The blue-shaded area shows the range of IGMF parameters that will be probed by the \gr\ observatory CTA \citep{Korochkin:2020pvg}.
    The red and black ticks over the BJ 04 and HS 22 recombination lines correspond to the range of magnetic field strengths obtained in \cite{Galli:2021mxk}, 
    which would induce enough baryon clumping to help 
   alleviate the Hubble tension, as proposed  in \cite{Jedamzik:2020krr}.
   The green and purple areas denote, respectively, the range of non-helical and helical magnetic field parameters that would arise from a first-order phase transition occurring at $T_* \sim 100$ GeV and sourcing a SGWB detectable by LISA,  fixing the smallest possible value $\varepsilon_{\rm turb}$ while still satisfying MAGIC's lower bound. These values of $\varepsilon_{\rm turb}$ are ${\cal O} (10^{-9})$ (non-helical) and ${\cal O} (10^{-13})$ (helical).}
    \label{fig:B_limits}
\end{figure*}

After its generation at the phase transition, the magnetic field evolves following the laws of turbulent decay up to the recombination epoch
(cf., ~\cite{Durrer:2013pga}).
Depending on the initial strength $B_*$, the initial correlation length $\lambda_*$, and the initial fractional helicity $h$, different turbulent evolutionary paths will lead to different IGMF today.  
In Fig.~\ref{fig:B_limits} we report, as initial conditions for the turbulent evolution, the regions in the parameter space given by the comoving initial strength $\tilde B$ and correlation length  $\tilde\lambda_B$ that are compatible
with a SGWB detectable by LISA,
according to the analyses presented in Sec.~\ref{results_analysis}
(Figs.~\ref{fig:alpha_beta_GW_signatures} and \ref{fig:GW_B_signatures}).
Since the resulting SGWB signal is expected to depend very weakly 
on the fractional helicity \citep{RoperPol:2021xnd}, the initial condition regions are considered to be valid both for non-helical and helical magnetic fields.
The black and gray-shaded areas correspond to
$\varepsilon_{\rm turb} = 1$ and $0.1$, respectively.
The orange-shaded region corresponds to the one found in \cite{RoperPol:2022iel}, where the limiting case was analysed, in which no sound waves are formed and all the kinetic energy from the plasma motion is in the form of MHD turbulence.
It is clear that including the sound-wave contribution to the SGWB spectrum enlarges the range of magnetic field parameters compatible with a detectable GW signal.

Examples of possible evolutionary paths for helical and non-helical fields are also shown in Fig.~\ref{fig:B_limits} by inclined lines starting at the boundaries of the initial condition regions, with arrows pointing at the parameter values
expected to be reached at recombination.
The evolutionary paths depend on the magnetic field helicity. 
Helical magnetic fields with maximal helicity $h=1$ are expected to follow the evolutionary path with $\tilde B\propto \tilde \lambda_B^{-1/2}$ in the parameter space $(\tilde B, \tilde \lambda_B)$, due to helicity conservation \citep[see dash-dotted purple lines in Fig.~\ref{fig:B_limits};][]{Banerjee:2004df,Durrer:2013pga}. 
The decay laws of non-helical fields are not entirely understood. 
The hypothesis of selective decay of short-range modes would lead to the decay law $\tilde B\propto \tilde \lambda_B^{-5/2}$, while the hypothesis of conservation of the Hosking invariant \citep{Hosking:2022umv,Zhou:2022xhk} suggests the decay law  $\tilde B\propto \tilde \lambda_B^{-5/4}$ (respectively, dashed and solid green lines in Fig.~\ref{fig:B_limits}). 

The endpoint of the evolution is also uncertain, depending on whether the evolution timescale is determined by the Alfv\'enic \citep{Banerjee:2004df} or the reconnection speed \citep{Hosking:2022umv}.
A straightforward estimate of the largest processed eddies at recombination suggests that the modes with $\tilde \lambda \lesssim \vA t_{\rm rec}$
are processed by turbulence
\citep{Banerjee:2004df}, with $\vA$ being the Alfv\'en velocity, defined in Eq.~(\ref{Alfven_fn}), and $t_{\rm rec}$ the Hubble time at recombination. 
In this case, setting $\tilde \lambda_{B} \equiv v_{\rm A} t_{\rm rec}$, the final strength and the correlation length satisfy the relation \citep{Durrer:2013pga}
\begin{equation}
        \tilde B \simeq 10^{-8}\left[\frac{\tilde \lambda_B}{1\mbox{ Mpc}}\right]
        \mbox{ G},
    \label{eq:endpoint}
\end{equation}
shown by the thick dash-dotted red line in Fig.~\ref{fig:B_limits}.
On the other hand, the evolution model put forward in  \cite{Hosking:2022umv}  proposes that the relevant timescale for the turbulent evolution is the one determined by reconnection, and that the decay law for non-helical fields is governed by the conservation of the Hosking integral, related to local fluctuations of helicity. 
Within this model, the correlation length at recombination can be smaller than the estimate in Eq.~(\ref{eq:endpoint}) by a factor of 20 for magnetic fields with final strength higher than $10^{-11}$~G, and even smaller for weaker fields—as much as by a factor of $10^{5.5}$ when the field strength is below $10^{-13}$ G \citep{Hosking:2022umv}. 
The possible endpoints of the reconnection-driven evolution are represented by the thick dash-dotted black line in Fig.~\ref{fig:B_limits}.

Figure~\ref{fig:B_limits} shows that the strength and correlation length of the IGMF resulting from the initial conditions leading to an SGWB detectable by LISA 
vary over a wide range:  $10^{-9}$~Mpc~$<\tilde \lambda_B<10^{-1}$~Mpc and $10^{-17}$~G~$<B<10^{-9}$~G,
depending on the fractional helicity $h$, on the uncertainty about 
the evolutionary path of non-helical fields, and on the uncertainty about the locus of the evolutionary endpoints.

It is also interesting to note that 
the phase transition parameters $(\alpha, \beta, v_w, T_*, \varepsilon_{\rm turb})$ providing a detectable SGWB correspond to a relic IGMF  that could be probed with \gr\  observations or through its imprint on recombination. 
Given the uncertainty of the magnetic field helicity $0<h<1$, none of the $(\alpha, \beta, v_w, T_*, \varepsilon_{\rm turb})$ combinations detectable by LISA are currently ruled out by known upper bounds on the magnetic field inferred from the CMB \citep{Planck:2015zrl}, Faraday rotation (FR) measures \citep{Pshirkov:2015tua}, or ultra-high-energy cosmic rays (UHECRs; \citealt{Neronov:2021xua}).
It has been proposed that primordial magnetic fields would also affect the CMB anisotropies through the production of small-scale baryonic density
fluctuations \citep{Jedamzik:2018itu}.
This leads to a stringent upper limit on the magnetic field amplitude, shown by the  upper ticks on the
solid red and black lines in Fig.~\ref{fig:B_limits} \citep{Jedamzik:2018itu,Galli:2021mxk}.
According to this upper limit, a small range of the parameters detectable by LISA (i.e., large $B_*$ and $\lambda_*$ or, equivalently, large $\alpha$, small $\beta/H_*$, and small $T_*$) would be ruled out if the magnetic field is helical.

Remarkably, within the evolution model of \citet{Hosking:2022umv}, all initial conditions corresponding to a detectable SGWB would also result in an IGMF consistent with the lower bounds of the MAGIC \gr\ observations \citep{Neronov:2010gir,MAGIC:2022piy}.
Moreover, if the magnetic field is non-helical, the resulting field strength and correlation length would be within the sensitivity reach of the next-generation \gr\ observatory CTA \citep{Korochkin:2020pvg}, again for all initial conditions compatible with a detectable SGWB. 
Only a present-day IGMF with strength  $\tilde B \gtrsim 10^{-11}$~G may not be accessible to CTA: this would correspond to  most of the initial condition parameter space if the magnetic field is helical. 
However, in this case, the field should have left an observable imprint at recombination, detectable in the CMB  \citep{Jedamzik:2018itu,Galli:2021mxk}. 
In any case, within the evolution model of \citet{Hosking:2022umv}, first-order phase transitions providing a detectable SGWB would have a complementary multi-messenger signature, either in the \gr\ or in the CMB data, even if the fraction of kinetic energy transferred to MHD turbulence is minimal.

This does not occur, however, within the evolution model of \citet{Banerjee:2004df}.
This model proposes that the Alfv\'en speed determines the relevant timescale for the turbulent evolution, and that non-helical fields decay following the hypothesis of selective decay of short-range modes. 
Following this evolution, a non-helical magnetic field could have a present-day strength that is below the \gr\ lower bound. 
In this case, \gr\ observations would not rule out the corresponding $(\alpha, \beta, v_w, T_*, \varepsilon_{\rm turb})$ combination; they just imply that
the observed IGMF must have been produced by a different process and that the field generated by the phase transition is not detectable on top of the dominant IGMF.

The range of IGMF parameters that correspond to a field strength in recombination that could help alleviate the Hubble tension via baryon clumping \citep{Jedamzik:2020krr,Galli:2021mxk}, is particularly interesting.
It is indicated by the GPJB 22 label and the red interval in Fig.~\ref{fig:B_limits}, if one assumes that the Alfv\'enic speed determines the final values at recombination \citep{Banerjee:2004df}; and by the black interval if one considers the relevant timescale to be that of reconnection \citep{Hosking:2022umv}. 
We find that for almost any beyond-the-Standard-Model scenario resulting in a first-order phase transition that produces an SGWB detectable by LISA, there exists some fractional helicity $h$, for which the magnetic field generated at the phase transition evolves into a sufficient strength at recombination to induce enough baryon clumping and to help alleviate  the Hubble tension. 
In this case, by combining a possible estimation of the magnetic field at recombination inferred from CMB data, with the other possible estimations inferred from the detection of a SGWB with LISA, one could fix the start and end points of the evolutionary path of the magnetic field and, in this way, provide a measurement of the initial helicity of the field. 

Finally, note that the transfer from the sound-wave kinetic energy to MHD turbulence does not need to be very efficient to guarantee the multi-messenger signature. 
Even if almost all of the energy in sound waves dissipated before leading
to the development of nonlinearities and MHD turbulence,
i.e.,  $\varepsilon_{\rm turb} \ll 1$, then the SGWB would simply be dominated by the sound-wave contribution, and it would still be detectable for appreciable ranges of the phase transition parameters $(\alpha, \beta, v_w, T_*)$. 
Correspondingly, even if $\varepsilon_{\rm turb} \ll 1$, there may still exist regions in the magnetic field parameter space that are compatible with the lower bounds of \gr\ observations. 
Fixing $T_* = 100$ GeV, we find that this happens provided that $\varepsilon_{\rm turb} \gtrsim {\cal O} (10^{-13})$ for non-helical magnetic fields, and  $\varepsilon_{\rm turb} \gtrsim {\cal O} (10^{-9})$ for helical fields. 
These initial conditions are shown, respectively, by the green and purple regions in Fig.~\ref{fig:B_limits}.

\section{Conclusions}

We analysed the possibility of obtaining a multi-messenger signature of a cosmological first-order phase transition if it leads to both a stochastic gravitational wave background detectable by LISA, and the generation of a primordial magnetic field that: i) leaves an  imprint at recombination observable in the CMB data, and ii) populates the voids of the LSS at present time with a strong enough intergalactic magnetic field to satisfy the lower bounds from \gr\ observations.
This multi-messenger signature could be present even  if a tiny fraction of the sound-wave kinetic energy is transferred to MHD turbulence. 

Adopting a phenomenological description of the phase transition in terms of five parameters $(\alpha, \beta, v_w, T_*, \varepsilon_{\rm turb})$, we found the ranges of these parameters for which the gravitational wave background is within LISA's sensitivity reach (Fig.~\ref{fig:alpha_beta_GW_signatures}). 
For each of such parameter combinations, we estimated the initial strength and correlation length of the primordial magnetic field that could be generated at the phase transition (Fig.~\ref{fig:GW_B_signatures}). From these initial conditions, we traced the field evolution in time until the present-day Universe, depending on the initial fractional helicity of the field (Fig.~\ref{fig:B_limits}). 
To do so, we adopted two possible evolution scenarios, proposed in  \citet{Banerjee:2004df} and in \citet{Hosking:2022umv}. 
We find that in the latter case, all first-order phase transition parameter combinations that provide a gravitational wave background detectable by LISA  result in a magnetic field that would be detectable either via \gr\ observations, for example, with the CTA telescope, or through its imprint on the CMB. 
If the cosmological evolution of the magnetic field follows the model of \citet{Banerjee:2004df}, it may happen that the resultant magnetic field
from the phase transition is too weak to explain the intergalactic magnetic field.

\section*{Data availability}

The templates of the different contributions (i.e., sound waves and turbulence) to the SGWB, the analysis tools and
results of this work (accompanied by a tutorial), and the different constraints on the IGMF, are available as part
of the open-source, public Python package {{\sc CosmoGW}}, stored on GitHub
\citep{CosmoGW_GH}.
The tutorial is available \href{https://github.com/CosmoGW/CosmoGW/blob/main/tutorials/GWs_MF_from_FOPT.ipynb}{here}.

\begin{acknowledgements}
    Support through the French
        National Research Agency (ANR) project MMUniverse (ANR-19-CE31-0020) is gratefully acknowledged.
       A.R.P. acknowledges support by the Swiss National Science Foundation (SNSF Ambizione grant \href{https://data.snf.ch/grants/grant/208807}{182044}).
    C.C. was supported by the Swiss National
    Science Foundation (SNSF Project Funding grant \href{https://data.snf.ch/grants/grant/212125}{212125})
    during the development of this project.
\end{acknowledgements}

\bibliographystyle{aa}
\bibliography{refs}

@article{Caprini:2015zlo,
    author = "Caprini, Chiara and others",
    title = "{Science with the space-based interferometer eLISA. II: Gravitational waves from cosmological phase transitions}",
    eprint = "1512.06239",
    archivePrefix = "arXiv",
    primaryClass = "astro-ph.CO",
    reportNumber = "DESY-15-246",
    doi = "10.1088/1475-7516/2016/04/001",
    journal = "JCAP",
    volume = "04",
    pages = "001",
    year = "2016"
}

@article{Cutting:2019zws,
    author = "Cutting, Daniel and Hindmarsh, Mark and Weir, David J.",
    title = "{Vorticity, kinetic energy, and suppressed gravitational wave production in strong first order phase transitions}",
    eprint = "1906.00480",
    archivePrefix = "arXiv",
    primaryClass = "hep-ph",
    reportNumber = "HIP-2019-15/TH",
    doi = "10.1103/PhysRevLett.125.021302",
    journal = "Phys. Rev. Lett.",
    volume = "125",
    number = "2",
    pages = "021302",
    year = "2020"
}

@article{Hindmarsh:2013xza,
    author = "Hindmarsh, Mark and Huber, Stephan J. and Rummukainen, Kari and Weir, David J.",
    title = "{Gravitational waves from the sound of a first order phase transition}",
    eprint = "1304.2433",
    archivePrefix = "arXiv",
    primaryClass = "hep-ph",
    reportNumber = "HIP-2013-07-TH",
    doi = "10.1103/PhysRevLett.112.041301",
    journal = "Phys. Rev. Lett.",
    volume = "112",
    pages = "041301",
    year = "2014"
}

@article{LISA:2024hlh,
    author = "Colpi, Monica and others",
    collaboration = "LISA",
    title = "{LISA Definition Study Report}",
    eprint = "2402.07571",
    archivePrefix = "arXiv",
    primaryClass = "astro-ph.CO",
    month = "2",
    year = "2024"
}

@article{PencilCode:2020eyn,
    author = "Brandenburg, A. and others",
    collaboration = "Pencil Code",
    title = "{The Pencil Code, a modular MPI code for partial differential equations and particles: multipurpose and multiuser-maintained}",
    eprint = "2009.08231",
    archivePrefix = "arXiv",
    primaryClass = "astro-ph.IM",
    reportNumber = "NORDITA-2020-087",
    doi = "10.21105/joss.02807",
    journal = "J. Open Source Softw.",
    volume = "6",
    number = "58",
    pages = "2807",
    year = "2021"
}

@article{RoperPol:2025lgc,
    author = "Roper Pol, Alberto and Midiri, Antonino Salvino",
    title = "{Relativistic magnetohydrodynamics in the early Universe}",
    eprint = "2501.05732",
    archivePrefix = "arXiv",
    journal = "JCAP (in press)",
    primaryClass = "gr-qc",
    month = "1",
    year = "2025"
}

@article{Correia:2025qif,
    author = "Correia, Jos{\'e} and Hindmarsh, Mark and Rummukainen, Kari and Weir, David J.",
    title = "{Gravitational waves from strong first-order phase transitions}",
    eprint = "2505.17824",
    archivePrefix = "arXiv",
    primaryClass = "astro-ph.CO",
    doi = "10.1103/8wmq-f635",
    journal = "Phys. Rev. D",
    volume = "112",
    number = "12",
    pages = "123546",
    year = "2025"
}

@article{Caprini:2024gyk,
    author = "Caprini, Chiara and Jinno, Ryusuke and Konstandin, Thomas and Roper Pol, Alberto and Rubira, Henrique and Stomberg, Isak",
    title = "{Gravitational waves from first-order phase transitions: from weak to strong}",
    eprint = "2409.03651",
    archivePrefix = "arXiv",
    primaryClass = "gr-qc",
    doi = "10.1007/JHEP07(2025)217",
    journal = "JHEP",
    volume = "07",
    pages = "217",
    year = "2025"
}

@article{Caprini:2024hue,
    author = "Caprini, Chiara and Jinno, Ryusuke and Lewicki, Marek and Madge, Eric and Merchand, Marco and Nardini, Germano and Pieroni, Mauro and Roper Pol, Alberto and Vaskonen, Ville",
    collaboration = "LISA Cosmology Working Group",
    title = "{Gravitational waves from first-order phase transitions in LISA: reconstruction pipeline and physics interpretation}",
    eprint = "2403.03723",
    archivePrefix = "arXiv",
    primaryClass = "astro-ph.CO",
    reportNumber = "LISA-COSWG-24-01, CERN-TH-2024-029",
    doi = "10.1088/1475-7516/2024/10/020",
    journal = "JCAP",
    volume = "10",
    pages = "020",
    year = "2024"
}

@article{Cai:2023guc,
    author = "Cai, Rong-Gen and Wang, Shao-Jiang and Yuwen, Zi-Yan",
    title = "{Hydrodynamic sound shell model}",
    eprint = "2305.00074",
    archivePrefix = "arXiv",
    primaryClass = "gr-qc",
    doi = "10.1103/PhysRevD.108.L021502",
    journal = "Phys. Rev. D",
    volume = "108",
    number = "2",
    pages = "L021502",
    year = "2023"
}

@article{Gowling:2021gcy,
    author = "Gowling, Chloe and Hindmarsh, Mark",
    title = "{Observational prospects for phase transitions at LISA: Fisher matrix analysis}",
    eprint = "2106.05984",
    archivePrefix = "arXiv",
    primaryClass = "astro-ph.CO",
    doi = "10.1088/1475-7516/2021/10/039",
    journal = "JCAP",
    volume = "10",
    pages = "039",
    year = "2021"
}

@ARTICLE{2011ApJ...727L...4D,
       author = {{Dolag}, K. and {Kachelriess}, M. and {Ostapchenko}, S. and {Tom{\`a}s}, R.},
        title = "{Lower Limit on the Strength and Filling Factor of Extragalactic Magnetic Fields}",
      journal = {\apjl},
         year = 2011,
        month = jan,
       volume = {727},
       number = {1},
          eid = {L4},
        pages = {L4},
          doi = {10.1088/2041-8205/727/1/L4},
archivePrefix = {arXiv},
       eprint = {1009.1782},
 primaryClass = {astro-ph.HE},
       adsurl = {https://ui.adsabs.harvard.edu/abs/2011ApJ...727L...4D}
}

@article{Hindmarsh:2019phv,
    author = "Hindmarsh, Mark and Hijazi, Mulham",
    title = "{Gravitational waves from first order cosmological phase transitions in the Sound Shell Model}",
    eprint = "1909.10040",
    archivePrefix = "arXiv",
    primaryClass = "astro-ph.CO",
    reportNumber = "NORDITA-2019-083, HIP-2019-29/TH",
    doi = "10.1088/1475-7516/2019/12/062",
    journal = "JCAP",
    volume = "12",
    pages = "062",
    year = "2019"
}

@article{Bodeker:2017cim,
    author = "Bodeker, Dietrich and Moore, Guy D.",
    title = "{Electroweak Bubble Wall Speed Limit}",
    eprint = "1703.08215",
    archivePrefix = "arXiv",
    primaryClass = "hep-ph",
    doi = "10.1088/1475-7516/2017/05/025",
    journal = "JCAP",
    volume = "05",
    pages = "025",
    year = "2017"
}

@article{Pshirkov:2015tua,
    author = "Pshirkov, M. S. and Tinyakov, P. G. and Urban, F. R.",
    title = "{New limits on extragalactic magnetic fields from rotation measures}",
    eprint = "1504.06546",
    archivePrefix = "arXiv",
    primaryClass = "astro-ph.CO",
    doi = "10.1103/PhysRevLett.116.191302",
    journal = "Phys. Rev. Lett.",
    volume = "116",
    number = "19",
    pages = "191302",
    year = "2016"
}

@article{Kahniashvili:2009qi,
    author = "Kahniashvili, Tina and Tevzadze, Alexander G. and Ratra, Bharat",
    title = "{Phase transition generated cosmological magnetic field at large scales}",
    eprint = "0907.0197",
    archivePrefix = "arXiv",
    primaryClass = "astro-ph.CO",
    doi = "10.1088/0004-637X/726/2/78",
    journal = "Astrophys. J.",
    volume = "726",
    pages = "78",
    year = "2011"
}

@article{Caprini:2019egz,
    author = "Caprini, Chiara and others",
    title = "{Detecting gravitational waves from cosmological phase transitions with LISA: an update}",
    eprint = "1910.13125",
    archivePrefix = "arXiv",
    primaryClass = "astro-ph.CO",
    reportNumber = "DESY-19-159, IPPP/19/27, HIP-2019-14/TH, MITP/19-066, IFT-UAM/CSIC-19-139",
    doi = "10.1088/1475-7516/2020/03/024",
    journal = "JCAP",
    volume = "03",
    pages = "024",
    year = "2020"
}

@inproceedings{Stomberg:2025kxf,
    author = "Stomberg, Isak and Roper Pol, Alberto",
    title = "{Gravitational wave spectra for cosmological phase transitions with non-linear decay of the fluid motion}",
    booktitle = "{Proceedings of the 59th Rencontres de Moriond on Gravitation}: {Moriond 2025 Gravitation}",
    eprint = "2508.04263",
    archivePrefix = "arXiv",
    primaryClass = "gr-qc",
    month = "8",
    year = "2025"
}

@article{Durrer:2013pga,
    author = "Durrer, Ruth and Neronov, Andrii",
    title = "{Cosmological Magnetic Fields: Their Generation, Evolution and Observation}",
    eprint = "1303.7121",
    archivePrefix = "arXiv",
    primaryClass = "astro-ph.CO",
    doi = "10.1007/s00159-013-0062-7",
    journal = "Astron. Astrophys. Rev.",
    volume = "21",
    pages = "62",
    year = "2013"
}

@article{Korochkin:2020pvg,
    author = "Korochkin, Alexander and Kalashev, Oleg and Neronov, Andrii and Semikoz, Dmitry",
    title = "{Sensitivity reach of gamma-ray measurements for strong cosmological magnetic fields}",
    eprint = "2007.14331",
    archivePrefix = "arXiv",
    primaryClass = "astro-ph.CO",
    doi = "10.3847/1538-4357/abc697",
    journal = "Astrophys. J.",
    volume = "906",
    number = "2",
    pages = "116",
    year = "2021"
}

@article{Neronov:2021xua,
    author = "Neronov, Andrii and Semikoz, Dmitri and Kalashev, Oleg",
    title = "{Limit on the intergalactic magnetic field from the ultrahigh-energy cosmic ray hotspot in the Perseus-Pisces region}",
    eprint = "2112.08202",
    archivePrefix = "arXiv",
    primaryClass = "astro-ph.HE",
    reportNumber = "INR-TH-2021-027",
    doi = "10.1103/PhysRevD.108.103008",
    journal = "Phys. Rev. D",
    volume = "108",
    number = "10",
    pages = "103008",
    year = "2023"
}

@article{Neronov:2010gir,
    author = "Neronov, A. and Vovk, I.",
    title = "{Evidence for strong extragalactic magnetic fields from Fermi observations of TeV blazars}",
    eprint = "1006.3504",
    archivePrefix = "arXiv",
    primaryClass = "astro-ph.HE",
    doi = "10.1126/science.1184192",
    journal = "Science",
    volume = "328",
    pages = "73--75",
    year = "2010"
}

@article{Planck:2015zrl,
    author = "Ade, P. A. R. and others",
    collaboration = "Planck",
    title = "{Planck 2015 results. XIX. Constraints on primordial magnetic fields}",
    eprint = "1502.01594",
    archivePrefix = "arXiv",
    primaryClass = "astro-ph.CO",
    doi = "10.1051/0004-6361/201525821",
    journal = "Astron. Astrophys.",
    volume = "594",
    pages = "A19",
    year = "2016"
}

@article{EPTA:2023fyk,
    author = "Antoniadis, J. and others",
    collaboration = "EPTA, InPTA:",
    title = "{The second data release from the European Pulsar Timing Array - III. Search for gravitational wave signals}",
    eprint = "2306.16214",
    archivePrefix = "arXiv",
    primaryClass = "astro-ph.HE",
    doi = "10.1051/0004-6361/202346844",
    journal = "Astron. Astrophys.",
    volume = "678",
    pages = "A50",
    year = "2023"
}

@article{
doi:10.1073/pnas.34.11.530,
author = {Theodore von Kármán },
title = {Progress in the Statistical Theory of Turbulence*},
journal = {Proceedings of the National Academy of Sciences},
volume = {34},
number = {11},
pages = {530-539},
year = {1948},
doi = {10.1073/pnas.34.11.530},
URL = {https://www.pnas.org/doi/abs/10.1073/pnas.34.11.530},
eprint = {https://www.pnas.org/doi/pdf/10.1073/pnas.34.11.530}}

@article{Vachaspati:1991nm,
    author = "Vachaspati, T.",
    title = "{Magnetic fields from cosmological phase transitions}",
    doi = "10.1016/0370-2693(91)90051-Q",
    journal = "Phys. Lett. B",
    volume = "265",
    pages = "258--261",
    year = "1991"
}

@article{Vachaspati:2001nb,
    author = "Vachaspati, Tanmay",
    title = "{Estimate of the primordial magnetic field helicity}",
    eprint = "astro-ph/0101261",
    archivePrefix = "arXiv",
    doi = "10.1103/PhysRevLett.87.251302",
    journal = "Phys. Rev. Lett.",
    volume = "87",
    pages = "251302",
    year = "2001"
}

@article{Xu:2023wog,
    author = "Xu, Heng and others",
    title = "{Searching for the Nano-Hertz Stochastic Gravitational Wave Background with the Chinese Pulsar Timing Array Data Release I}",
    eprint = "2306.16216",
    archivePrefix = "arXiv",
    primaryClass = "astro-ph.HE",
    doi = "10.1088/1674-4527/acdfa5",
    journal = "Res. Astron. Astrophys.",
    volume = "23",
    number = "7",
    pages = "075024",
    year = "2023"
}

@article{Sharma:2023mao,
    author = "Sharma, Ramkishor and Dahl, Jani and Brandenburg, Axel and Hindmarsh, Mark",
    title = "{Shallow relic gravitational wave spectrum with acoustic peak}",
    eprint = "2308.12916",
    archivePrefix = "arXiv",
    primaryClass = "gr-qc",
    reportNumber = "NORDITA-2023-051, HIP-2023-13/TH",
    doi = "10.1088/1475-7516/2023/12/042",
    journal = "JCAP",
    volume = "12",
    pages = "042",
    year = "2023"
}

@article{RoperPol:2023dzg,
    author = "Roper Pol, Alberto and Procacci, Simona and Caprini, Chiara",
    title = "{Characterization of the gravitational wave spectrum from sound waves within the sound shell model}",
    eprint = "2308.12943",
    archivePrefix = "arXiv",
    primaryClass = "gr-qc",
    doi = "10.1103/PhysRevD.109.063531",
    journal = "Phys. Rev. D",
    volume = "109",
    number = "6",
    pages = "063531",
    year = "2024"
}

@article{Miniati:2017kah,
    author = "Miniati, Francesco and Gregori, Gianluca and Reville, Brian and Sarkar, Subir",
    title = "{Axion-Driven Cosmic Magnetogenesis during the QCD Crossover}",
    eprint = "1708.07614",
    archivePrefix = "arXiv",
    primaryClass = "astro-ph.CO",
    doi = "10.1103/PhysRevLett.121.021301",
    journal = "Phys. Rev. Lett.",
    volume = "121",
    number = "2",
    pages = "021301",
    year = "2018"
}

@article{Zhang:2019vsb,
    author = "Zhang, Yiyang and Vachaspati, Tanmay and Ferrer, Francesc",
    title = "{Magnetic field production at a first-order electroweak phase transition}",
    eprint = "1902.02751",
    archivePrefix = "arXiv",
    primaryClass = "hep-ph",
    doi = "10.1103/PhysRevD.100.083006",
    journal = "Phys. Rev. D",
    volume = "100",
    number = "8",
    pages = "083006",
    year = "2019"
}

@article{Reardon:2023gzh,
    author = "Reardon, Daniel J. and others",
    title = "{Search for an Isotropic Gravitational-wave Background with the Parkes Pulsar Timing Array}",
    eprint = "2306.16215",
    archivePrefix = "arXiv",
    primaryClass = "astro-ph.HE",
    doi = "10.3847/2041-8213/acdd02",
    journal = "Astrophys. J. Lett.",
    volume = "951",
    number = "1",
    pages = "L6",
    year = "2023"
}

@article{NANOGrav:2023gor,
    author = "Agazie, Gabriella and others",
    collaboration = "NANOGrav",
    title = "{The NANOGrav 15 yr Data Set: Evidence for a Gravitational-wave Background}",
    eprint = "2306.16213",
    archivePrefix = "arXiv",
    primaryClass = "astro-ph.HE",
    doi = "10.3847/2041-8213/acdac6",
    journal = "Astrophys. J. Lett.",
    volume = "951",
    number = "1",
    pages = "L8",
    year = "2023"
}

@article{Durrer:2003ja,
    author = "Durrer, Ruth and Caprini, Chiara",
    title = "{Primordial magnetic fields and causality}",
    eprint = "astro-ph/0305059",
    archivePrefix = "arXiv",
    doi = "10.1088/1475-7516/2003/11/010",
    journal = "JCAP",
    volume = "11",
    pages = "010",
    year = "2003"
}

@article{Brandenburg:2021tmp,
    author = "Brandenburg, Axel and Clarke, Emma and He, Yutong and Kahniashvili, Tina",
    title = "{Can we observe the QCD phase transition-generated gravitational waves through pulsar timing arrays?}",
    eprint = "2102.12428",
    archivePrefix = "arXiv",
    primaryClass = "astro-ph.CO",
    reportNumber = "NORDITA-2021-016",
    doi = "10.1103/PhysRevD.104.043513",
    journal = "Phys. Rev. D",
    volume = "104",
    number = "4",
    pages = "043513",
    year = "2021"
}

@article{RoperPol:2021xnd,
    author = "Roper Pol, Alberto and Mandal, Sayan and Brandenburg, Axel and Kahniashvili, Tina",
    title = "{Polarization of gravitational waves from helical MHD turbulent sources}",
    eprint = "2107.05356",
    archivePrefix = "arXiv",
    primaryClass = "gr-qc",
    reportNumber = "NORDITA-2021-062",
    doi = "10.1088/1475-7516/2022/04/019",
    journal = "JCAP",
    volume = "04",
    number = "04",
    pages = "019",
    year = "2022"
}

@article{Subramanian:2015lua,
    author = "Subramanian, Kandaswamy",
    title = "{The origin, evolution and signatures of primordial magnetic fields}",
    eprint = "1504.02311",
    archivePrefix = "arXiv",
    primaryClass = "astro-ph.CO",
    doi = "10.1088/0034-4885/79/7/076901",
    journal = "Rept. Prog. Phys.",
    volume = "79",
    number = "7",
    pages = "076901",
    year = "2016"
}

@article{NANOGrav:2023hvm,
    author = "Afzal, Adeela and others",
    collaboration = "NANOGrav",
    title = "{The NANOGrav 15 yr Data Set: Search for Signals from New Physics}",
    eprint = "2306.16219",
    archivePrefix = "arXiv",
    primaryClass = "astro-ph.HE",
    doi = "10.3847/2041-8213/acdc91",
    journal = "Astrophys. J. Lett.",
    volume = "951",
    number = "1",
    pages = "L11",
    year = "2023"
}

@article{EPTA:2023xxk,
    author = "Antoniadis, J. and others",
    collaboration = "EPTA, InPTA",
    title = "{The second data release from the European Pulsar Timing Array - IV. Implications for massive black holes, dark matter, and the early Universe}",
    eprint = "2306.16227",
    archivePrefix = "arXiv",
    primaryClass = "astro-ph.CO",
    doi = "10.1051/0004-6361/202347433",
    journal = "Astron. Astrophys.",
    volume = "685",
    pages = "A94",
    year = "2024"
}

@article{Jedamzik:2018itu,
    author = "Jedamzik, Karsten and Saveliev, Andrey",
    title = "{Stringent Limit on Primordial Magnetic Fields from the Cosmic Microwave Background Radiation}",
    eprint = "1804.06115",
    archivePrefix = "arXiv",
    primaryClass = "astro-ph.CO",
    doi = "10.1103/PhysRevLett.123.021301",
    journal = "Phys. Rev. Lett.",
    volume = "123",
    number = "2",
    pages = "021301",
    year = "2019"
}

@article{Kahniashvili:2021gym,
    author = "Kahniashvili, Tina and Clarke, Emma and Stepp, Jonathan and Brandenburg, Axel",
    title = "{Big Bang Nucleosynthesis Limits and Relic Gravitational-Wave Detection Prospects}",
    eprint = "2111.09541",
    archivePrefix = "arXiv",
    primaryClass = "astro-ph.CO",
    reportNumber = "NORDITA-2021-089",
    doi = "10.1103/PhysRevLett.128.221301",
    journal = "Phys. Rev. Lett.",
    volume = "128",
    number = "22",
    pages = "221301",
    year = "2022"
}

@article{LISA:2017pwj,
    author = "Amaro-Seoane, Pau and others",
    collaboration = "LISA",
    title = "{Laser Interferometer Space Antenna}",
    eprint = "1702.00786",
    archivePrefix = "arXiv",
    primaryClass = "astro-ph.IM",
    month = "2",
    year = "2017"
}

@article{Brandenburg:2014mwa,
    author = "Brandenburg, Axel and Kahniashvili, Tina and Tevzadze, Alexander G.",
    title = "{Nonhelical inverse transfer of a decaying turbulent magnetic field}",
    eprint = "1404.2238",
    archivePrefix = "arXiv",
    primaryClass = "astro-ph.CO",
    reportNumber = "NORDITA-2014-42",
    doi = "10.1103/PhysRevLett.114.075001",
    journal = "Phys. Rev. Lett.",
    volume = "114",
    number = "7",
    pages = "075001",
    year = "2015"
}

@article{Brandenburg:1996fc,
    author = "Brandenburg, Axel and Enqvist, Kari and Olesen, Poul",
    title = "{Large scale magnetic fields from hydromagnetic turbulence in the very early universe}",
    eprint = "astro-ph/9602031",
    archivePrefix = "arXiv",
    reportNumber = "NORDITA-96-6-A",
    doi = "10.1103/PhysRevD.54.1291",
    journal = "Phys. Rev. D",
    volume = "54",
    pages = "1291--1300",
    year = "1996"
}

@article{Caprini:2018mtu,
    author = "Caprini, Chiara and Figueroa, Daniel G.",
    title = "{Cosmological Backgrounds of Gravitational Waves}",
    eprint = "1801.04268",
    archivePrefix = "arXiv",
    primaryClass = "astro-ph.CO",
    doi = "10.1088/1361-6382/aac608",
    journal = "Class. Quant. Grav.",
    volume = "35",
    number = "16",
    pages = "163001",
    year = "2018"
}

@article{RoperPol:2019wvy,
    author = "Roper Pol, Alberto and Mandal, Sayan and Brandenburg, Axel and Kahniashvili, Tina and Kosowsky, Arthur",
    title = "{Numerical simulations of gravitational waves from early-universe turbulence}",
    eprint = "1903.08585",
    archivePrefix = "arXiv",
    primaryClass = "astro-ph.CO",
    reportNumber = "NORDITA-2019-024",
    doi = "10.1103/PhysRevD.102.083512",
    journal = "Phys. Rev. D",
    volume = "102",
    number = "8",
    pages = "083512",
    year = "2020"
}

@article{Kamionkowski:1993fg,
    author = "Kamionkowski, Marc and Kosowsky, Arthur and Turner, Michael S.",
    title = "{Gravitational radiation from first order phase transitions}",
    eprint = "astro-ph/9310044",
    archivePrefix = "arXiv",
    reportNumber = "IASSNS-HEP-93-44, FERMILAB-PUB-93-235-A",
    doi = "10.1103/PhysRevD.49.2837",
    journal = "Phys. Rev. D",
    volume = "49",
    pages = "2837--2851",
    year = "1994"
}

@article{Stephanov:2006zvm,
    author = "Stephanov, M. A.",
    editor = "Blum, Tom and Creutz, Michael and DeTar, Carleton and Karsch, Frithjof and Kronfeld, Andreas and Morningstar, Colin and Richards, David and Shigemitsu, Junko and Toussaint, Doug",
    title = "{QCD phase diagram: An Overview}",
    eprint = "hep-lat/0701002",
    archivePrefix = "arXiv",
    doi = "10.22323/1.032.0024",
    journal = "PoS",
    volume = "LAT2006",
    pages = "024",
    year = "2006"
}

@article{Kajantie:1995kf,
    author = "Kajantie, K. and Laine, M. and Rummukainen, K. and Shaposhnikov, Mikhail E.",
    title = "{The Electroweak phase transition: A Nonperturbative analysis}",
    eprint = "hep-lat/9510020",
    archivePrefix = "arXiv",
    reportNumber = "CERN-TH-95-263, HD-THEP-95-44, HU-TFT-95-57, IUHET-318",
    doi = "10.1016/0550-3213(96)00052-1",
    journal = "Nucl. Phys. B",
    volume = "466",
    pages = "189--258",
    year = "1996"
}

@article{Vachaspati:2020blt,
    author = "Vachaspati, Tanmay",
    title = "{Progress on cosmological magnetic fields}",
    eprint = "2010.10525",
    archivePrefix = "arXiv",
    primaryClass = "astro-ph.CO",
    doi = "10.1088/1361-6633/ac03a9",
    journal = "Rept. Prog. Phys.",
    volume = "84",
    number = "7",
    pages = "074901",
    year = "2021"
}

@article{MAGIC:2022piy,
    author = "Acciari, V. A. and others",
    collaboration = "MAGIC",
    title = "{A lower bound on intergalactic magnetic fields from time variability of 1ES 0229+200 from MAGIC and Fermi/LAT observations}",
    eprint = "2210.03321",
    archivePrefix = "arXiv",
    primaryClass = "astro-ph.HE",
    doi = "10.1051/0004-6361/202244126",
    journal = "Astron. Astrophys.",
    volume = "670",
    pages = "A145",
    year = "2023"
}

@article{Shvartsman:1969mm,
    author = "Shvartsman, V. F.",
    title = "{Density of relict particles with zero rest mass in the universe}",
    journal = "Pisma Zh. Eksp. Teor. Fiz.",
    volume = "9",
    pages = "315--317",
    year = "1969"
}

@article{Ellis:2020awk,
    author = "Ellis, John and Lewicki, Marek and No, Jos\'e Miguel",
    title = "{Gravitational waves from first-order cosmological phase transitions: lifetime of the sound wave source}",
    eprint = "2003.07360",
    archivePrefix = "arXiv",
    primaryClass = "hep-ph",
    reportNumber = "KCL-PH-TH/2020-04, CERN-TH-2020-016, IFT-UAM/CSIC-20-35",
    doi = "10.1088/1475-7516/2020/07/050",
    journal = "JCAP",
    volume = "07",
    pages = "050",
    year = "2020"
}

@article{RoperPol:2022iel,
    author = "Roper Pol, Alberto and Caprini, Chiara and Neronov, Andrii and Semikoz, Dmitri",
    title = "{Gravitational wave signal from primordial magnetic fields in the Pulsar Timing Array frequency band}",
    eprint = "2201.05630",
    archivePrefix = "arXiv",
    primaryClass = "astro-ph.CO",
    doi = "10.1103/PhysRevD.105.123502",
    journal = "Phys. Rev. D",
    volume = "105",
    number = "12",
    pages = "123502",
    year = "2022"
}

@article{Jedamzik:2020krr,
    author = "Jedamzik, Karsten and Pogosian, Levon",
    title = "{Relieving the Hubble tension with primordial magnetic fields}",
    eprint = "2004.09487",
    archivePrefix = "arXiv",
    primaryClass = "astro-ph.CO",
    doi = "10.1103/PhysRevLett.125.181302",
    journal = "Phys. Rev. Lett.",
    volume = "125",
    number = "18",
    pages = "181302",
    year = "2020"
}

@article{Galli:2021mxk,
    author = "Galli, Silvia and Pogosian, Levon and Jedamzik, Karsten and Balkenhol, Lennart",
    title = "{Consistency of Planck, ACT, and SPT constraints on magnetically assisted recombination and forecasts for future experiments}",
    eprint = "2109.03816",
    archivePrefix = "arXiv",
    primaryClass = "astro-ph.CO",
    doi = "10.1103/PhysRevD.105.023513",
    journal = "Phys. Rev. D",
    volume = "105",
    number = "2",
    pages = "023513",
    year = "2022"
}

@article{Hosking:2022umv,
    author = "Hosking, David N. and Schekochihin, Alexander A.",
    title = "{Cosmic-void observations reconciled with primordial magnetogenesis}",
    eprint = "2203.03573",
    archivePrefix = "arXiv",
    primaryClass = "astro-ph.CO",
    doi = "10.1038/s41467-023-43258-3",
    journal = "Nature Commun.",
    volume = "14",
    number = "1",
    pages = "7523",
    year = "2023"
}

@article{Espinosa:2010hh,
    author = "Espinosa, Jose R. and Konstandin, Thomas and No, Jose M. and Servant, Geraldine",
    title = "{Energy Budget of Cosmological First-order Phase Transitions}",
    eprint = "1004.4187",
    archivePrefix = "arXiv",
    primaryClass = "hep-ph",
    reportNumber = "CERN-PH-TH-2010-027",
    doi = "10.1088/1475-7516/2010/06/028",
    journal = "JCAP",
    volume = "06",
    pages = "028",
    year = "2010"
}

@article{Banerjee:2004df,
    author = "Banerjee, Robi and Jedamzik, Karsten",
    title = "{The Evolution of cosmic magnetic fields: From the very early universe, to recombination, to the present}",
    eprint = "astro-ph/0410032",
    archivePrefix = "arXiv",
    doi = "10.1103/PhysRevD.70.123003",
    journal = "Phys. Rev. D",
    volume = "70",
    pages = "123003",
    year = "2004"
}

@article{Sigl:1996dm,
    author = "Sigl, Guenter and Olinto, Angela V. and Jedamzik, Karsten",
    title = "{Primordial magnetic fields from cosmological first order phase transitions}",
    eprint = "astro-ph/9610201",
    archivePrefix = "arXiv",
    doi = "10.1103/PhysRevD.55.4582",
    journal = "Phys. Rev. D",
    volume = "55",
    pages = "4582--4590",
    year = "1997"
}

@article{Auclair:2022jod,
    author = "Auclair, Pierre and Caprini, Chiara and Cutting, Daniel and Hindmarsh, Mark and Rummukainen, Kari and Steer, Dani\`ele A. and Weir, David J.",
    title = "{Generation of gravitational waves from freely decaying turbulence}",
    eprint = "2205.02588",
    archivePrefix = "arXiv",
    primaryClass = "astro-ph.CO",
    reportNumber = "HIP-2021-35/TH",
    doi = "10.1088/1475-7516/2022/09/029",
    journal = "JCAP",
    volume = "09",
    pages = "029",
    year = "2022"
}

@article{Kawasaki:2012va,
    author = "Kawasaki, Masahiro and Kusakabe, Motohiko",
    title = "{Updated constraint on a primordial magnetic field during big bang nucleosynthesis and a formulation of field effects}",
    eprint = "1204.6164",
    archivePrefix = "arXiv",
    primaryClass = "astro-ph.CO",
    reportNumber = "ICRR-REPORT-621-2012-10",
    doi = "10.1103/PhysRevD.86.063003",
    journal = "Phys. Rev. D",
    volume = "86",
    pages = "063003",
    year = "2012"
}

@article{Hindmarsh:2016lnk,
    author = "Hindmarsh, Mark",
    title = "{Sound shell model for acoustic gravitational wave production at a first-order phase transition in the early Universe}",
    eprint = "1608.04735",
    archivePrefix = "arXiv",
    primaryClass = "astro-ph.CO",
    doi = "10.1103/PhysRevLett.120.071301",
    journal = "Phys. Rev. Lett.",
    volume = "120",
    number = "7",
    pages = "071301",
    year = "2018"
}

@article{Hindmarsh:2017gnf,
    author = "Hindmarsh, Mark and Huber, Stephan J. and Rummukainen, Kari and Weir, David J.",
    title = "{Shape of the acoustic gravitational wave power spectrum from a first order phase transition}",
    eprint = "1704.05871",
    archivePrefix = "arXiv",
    primaryClass = "astro-ph.CO",
    reportNumber = "HIP-2017-02-TH, HIP-2017-02/TH",
    doi = "10.1103/PhysRevD.96.103520",
    journal = "Phys. Rev. D",
    volume = "96",
    number = "10",
    pages = "103520",
    year = "2017",
    note = "[Erratum: Phys.Rev.D 101, 089902 (2020)]"
}

@article{Hindmarsh:2020hop,
    author = {Hindmarsh, Mark B. and L\"uben, Marvin and Lumma, Johannes and Pauly, Martin},
    title = "{Phase transitions in the early universe}",
    eprint = "2008.09136",
    archivePrefix = "arXiv",
    primaryClass = "astro-ph.CO",
    reportNumber = "MPP-2020-163, HIP-2020-27/TH",
    doi = "10.21468/SciPostPhysLectNotes.24",
    journal = "SciPost Phys. Lect. Notes",
    volume = "24",
    pages = "1",
    year = "2021"
}

@article{Jinno:2022mie,
    author = "Jinno, Ryusuke and Konstandin, Thomas and Rubira, Henrique and Stomberg, Isak",
    title = "{Higgsless simulations of cosmological phase transitions and gravitational waves}",
    eprint = "2209.04369",
    archivePrefix = "arXiv",
    primaryClass = "astro-ph.CO",
    reportNumber = "DESY 22-148, IFT-UAM/CSIC-22-100, TUM-HEP-1416/22",
    doi = "10.1088/1475-7516/2023/02/011",
    journal = "JCAP",
    volume = "02",
    pages = "011",
    year = "2023"
}

@article{Zhou:2022xhk,
    author = "Zhou, Hongzhe and Sharma, Ramkishor and Brandenburg, Axel",
    title = "{Scaling of the Hosking integral in decaying magnetically dominated turbulence}",
    eprint = "2206.07513",
    archivePrefix = "arXiv",
    primaryClass = "physics.plasm-ph",
    reportNumber = "NORDITA 2022-040",
    doi = "10.1017/S002237782200109X",
    journal = "J. Plasma Phys.",
    volume = "88",
    number = "6",
    pages = "905880602",
    year = "2022"
}

@article{CosmoGW_GH,
    author = "Roper Pol, Alberto",
    doi = "10.5281/zenodo.6045844",
    title = "{{\sc CosmoGW} public repository, stored on \href{https://github.com/CosmoGW/CosmoGW}{GitHub};
    see documentation on \href{https://cosmogw-manual.readthedocs.io/en/latest/}{Read the Docs}}",
    journal = "{\sc CosmoGW} public repository, stored on \href{https://github.com/CosmoGW/CosmoGW}{GitHub};
    see documentation on \href{https://cosmogw-manual.readthedocs.io/en/latest/}{Read the Docs}.",
    year = "2024"
}
\end{document}